# Predictive Control based on Equivalent Dynamic Linearization Model

Feilong Zhang

***Abstract*—Based on equivalent-dynamic-linearization model (EDLM), we propose a kind of model predictive control (MPC) for single-input and single-output (SISO) nonlinear or linear systems. After compensating the EDLM with disturbance for multiple-input and multiple-output nonlinear or linear systems, the MPC compensated with disturbance is proposed to address the disturbance rejection problem. The system performance analysis results are much clear compared with the system stability analyses on MPC in current works. And this may help the engineers understand how to design, analyze and apply the controller in practical.**

## I. Introduction

Since 1970 various model-based control methods have surged for multivariable control of industrial processes. Simultaneously, the interest in MPC started to surge after Richalet had published the pioneering paper on MPHC (MAC) [1], [2] and Cutler had introduced the dynamic matrix control (DMC) [3], although their methods are not designed for unstable systems. MPC represents a kind of digital control algorithm, which utilizes an explicit process model to predict its future behaviors. At each sampling instant an MPC algorithm is designed to optimize the cost function which is associated with predicted system behaviors determined by a sequence of future control variables. Only the first control in this on-line calculated optimal sequence is applied in the system, and this calculation repeats at the subsequent sampling instant. MPC can now be found in a variety of industrial settings including chemical processes, automotive and even the MIT's cheetah 3.

MPC derived from the basic multivariable process control task which requires the operation of processes within predefined operating constraints [4], [5]. There are two main reasons for the popularity of MPC in industrial applications. One is that MPC is a class of model-based advanced control strategy, which may easily handle the problems of time-delays, non-minimum phase and unstable properties when the process model is built appropriately.

Manuscript received Jun 28, 2021. This work was supported in part by the xxxxxxxxxxxx.

Feilong Zhang is with the State Key Laboratory of Robotics, Shenyang Institute of Automation, Chinese Academy of Sciences, Shenyang 110016, China (e-mail: zhangfeiong@sia.cn).

Another one is attributed to the easy extension to multivariable processes and the constraint-handling capabilities. From academic view, the researches on self-tuning control stimulate the interest in MPC. The minimum variance control was proposed by Åström to minimize the performance index only concerning tracking performance at the future time $k+d$, where $d$ is the process dead-time [7]. Additionally to handle the non-minimum phase systems, the generalized minimum variance control and generalized predictive control (GPC) scheme [8]-[10] were proposed to minimize the quadratic performance index. The GPC design is based on the auto regressive moving average (CARMA) models or auto regressive integrated moving average (CARIMA) models which both are for the linear system description and naturally introduce an integrator in the control design process. Further, it is necessary to solve the Diophantine equation for GPC design process, and this leads to the non-intuitive system characteristics in the system performance analysis [8]-[12]. Similarly, MPC designed based on state-space is studied and analyzed in a considerable amount of works [13]-[17]. To theoretically guarantee the system performance or stability by Lyapunov stability analysis, although the physical meaning of original problem is clear, performance modification indices (such as adding terminal penalty term), or supplement for artificial constraints, including terminal state constraints and terminal set constraints are normally necessitated. Consequently, the designed control algorithms and stability analysis become more conservative. In addition, unclear physical meanings in current MPC algorithm make it difficult to relate with application practice [18].

During this decade, a large number of papers about model-free adaptive control (MFAC) have been published [19], [20]. The MFAC design relies on a kind of process model referred to as equivalent-dynamic-linearization model (EDLM). Although many conclusions in current works contradict with [21]-[25], the EDLM is still useful in our controller design. In this paper, a new kind of predictive control based on EDLM is proposed for both SISO systems and MIMO systems. It can be regarded as another form of "model-free adaptive predictive control" in [21]. For more its merits, please refer to [21]. Compared with current works of unconstrained MPC, the physical meaning of each part of controller is much clearer both in controller design and system characteristic analysis. Furthermore, one merit of proposed MPC is that the system equation of unconstrained predictive control and that of unconstrained non-predictive control are unified in our papers.

Some contributions of this paper are summarized as follows. i) We design the MPC based on EDLM for the SISO nonlinear system and firstly analyze the system performance by the closed-loop equation and steady-state analysis for the nonlinear system. The results are much clearer than the aforementioned works. ii) The EDLM is compensated with disturbance for the description of MIMO nonlinear system subject to disturbance, then the MPC compensated with disturbance is designed. Similarly, the system performance and the influence of disturbance are analyzed by the closed-loop system equation and show that the influence of the known disturbance will be theoretically removed when the system stability is guaranteed and $\boldsymbol{\lambda}=\mathbf{0}$.

## II. System Description and Predictive Control Design for SISO Systems

### A. System description for SISO nonlinear systems

In this section, the EDLM for SISO systems is given as a fundamental tool for the predictive control design.

The discrete-time SISO nonlinear system is considered as follow:

$$y(k+1)=f\,(y(k),\cdots,y(k-n_y),u(k),\cdots,u(k-n_u)) \quad (1)$$

where $f(\cdots)\in$ R is a nonlinear differentiable function, $n_u$+1, $n_y$+1 $\in$ $Z$ are the orders of input $u(k)$ and the output $y(k)$ of the system at time $k$, respectively.

Define $\boldsymbol{\varphi}(k)=[y(k),\cdots,y(k-n_y),u(k),\cdots,u(k-n_u)]$, then (1) can be written as

$$y(k+1)=f\,(\boldsymbol{\varphi}(k)) \quad (2)$$

*Theorem 1:* If $\Delta\boldsymbol{H}(k)\neq\mathbf{0}$, there must exist a time-varying vector $\phi_L(k)$ called PG vector and the system (1) can be transformed into the EDLM shown as follows

$$\Delta y(k+1)=\boldsymbol{\phi}_L^T(k)\Delta\boldsymbol{H}(k) \quad (3)$$

where

$\boldsymbol{\phi}_L(k)=\begin{bmatrix}\boldsymbol{\phi}_{Ly}(k)\\ \boldsymbol{\phi}_{Lu}(k)\end{bmatrix}=[\phi_1(k),\cdots,\phi_{Ly}(k),\phi_{Ly+1}(k),\cdots,\phi_{Ly+Lu}(k)]^T$ and $\Delta\boldsymbol{H}(k)=\begin{bmatrix}\Delta\boldsymbol{Y}_{Ly}^T(k) & \Delta\boldsymbol{U}_{Lu}^T(k)\end{bmatrix}^T$ is a vector that contains the incremental system output vector $\Delta\boldsymbol{Y}_{Ly}(k)=[\Delta y^T(k),\cdots,\Delta y^T(k-L_y+1)]^T$ and the incremental system input vector $\Delta\boldsymbol{U}_{Lu}(k)=[\Delta u^T(k),\cdots,\Delta u^T(k-L_u+1)]^T$. Two integers $1\leq L_y$ and $0\leq L_u$ are called the pseudo orders. In this paper, we only analyze the case $L_y=n_y+1$ and $L_u=n_u+1$.

*Proof:* Please refer to Appendix.

### B. Predictive model

We can rewrite (3) into (4).

$$\begin{aligned}\Delta\boldsymbol{x}(k+1)&=\boldsymbol{A}(k)\Delta\boldsymbol{x}(k)+\boldsymbol{B}(k)\Delta\boldsymbol{u}(k)\\ \Delta y(k+1)&=\boldsymbol{C}\Delta\boldsymbol{x}(k+1)\end{aligned} \quad (4)$$

where

$\boldsymbol{B}^T=\begin{bmatrix}\phi_{Ly+1}(k),0,\cdots,0,1,0,\cdots,0\end{bmatrix}_{1\times(Ly+Lu)}$, $\boldsymbol{C}=\begin{bmatrix}1,0,\cdots,0\end{bmatrix}$,

$$\Delta\boldsymbol{x}(k)=\begin{bmatrix}\Delta\boldsymbol{Y}_{Ly}(k)\\ \Delta\boldsymbol{U}_{Lu}(k-1)\end{bmatrix}=[\Delta y(k),\cdots,\Delta y(k-L_y+1),\Delta u(k-1),\cdots,\Delta u(k-L_u)]^T$$

$\boldsymbol{A}(k)=$

$$\begin{bmatrix}\phi_1(k) & \phi_2(k) & \cdots & \phi_{Ly-1}(k) & \phi_{Ly}(k) & \phi_{Ly+2}(k) & \cdots & \phi_{Ly+Lu}(k) & 0\\ 1 & 0 & \cdots & 0 & 0 & 0 & \cdots & 0 & 0\\ 0 & 1 & \cdots & 0 & 0 & 0 & \ddots & 0 & 0\\ \vdots & \vdots & \ddots & \vdots & \vdots & \vdots & \ddots & \vdots & \vdots\\ 0 & 0 & \cdots & 1 & 0 & 0 & \cdots & 0 & 0\\ 0 & 0 & \cdots & 0 & 0 & 0 & \cdots & 0 & 0\\ 0 & 0 & \cdots & 0 & 0 & 1 & \ddots & 0 & 0\\ \vdots & \vdots & \ddots & \vdots & \vdots & \vdots & \ddots & \vdots & \vdots\\ 0 & 0 & \cdots & 0 & 0 & 0 & \cdots & 1 & 0\end{bmatrix}$$

Then we have finite $N$ step forward prediction model as follow:

$$\begin{aligned}\Delta\boldsymbol{x}(k+1)&=\boldsymbol{A}(k)\Delta\boldsymbol{x}(k)+\boldsymbol{B}(k)\Delta u(k)\\ \Delta\boldsymbol{x}(k+2)&=\boldsymbol{A}(k+1)\Delta\boldsymbol{x}(k+1)+\boldsymbol{B}(k+1)\Delta u(k+1)\\ &=\boldsymbol{A}(k+1)\boldsymbol{A}(k)\Delta\boldsymbol{x}(k)+\boldsymbol{A}(k+1)\boldsymbol{B}(k)\Delta u(k)\\ &\quad+\boldsymbol{B}(k+1)\Delta u(k+1)\\ &\vdots\\ \Delta\boldsymbol{x}(k+N)&=\boldsymbol{A}(k+N-1)\Delta\boldsymbol{x}(k+N-1)+\boldsymbol{B}(k+N-1)\Delta u(k+N-1)\\ &=\prod_{i=0}^{N-1}\boldsymbol{A}(k+i)\Delta\boldsymbol{x}(k)+\prod_{i=1}^{N-1}\boldsymbol{A}(k+i)\boldsymbol{B}(k)\Delta u(k)\\ &\quad+\prod_{i=2}^{N-1}\boldsymbol{A}(k+i)\boldsymbol{B}(k+1)\Delta u(k+1)+\cdots\\ &\quad+\boldsymbol{B}(k+N-1)\Delta u(k+N-1)\end{aligned} \quad (5)$$

where $N$ is the predictive step length, $\Delta y(k+i)$ and $\Delta u(k+i)$ are the increment values of the predictive output and the predictive input in the future time $k+i$ ($i$=1,2,$\cdots$,$N$ ), respectively. Here, we define $\boldsymbol{Y}_N(k)$, $\Delta\boldsymbol{Y}_N(k+1)$, $\Delta\boldsymbol{U}_N(k)$, $\boldsymbol{\Psi}(k)$, $\tilde{\boldsymbol{\Psi}}(k)$, $\boldsymbol{\Phi}(k)$, $\tilde{\boldsymbol{\Phi}}(k)$ as follows:

$\boldsymbol{\Phi}(k)=$

$$\begin{bmatrix}\boldsymbol{CB}(k) & 0 & \cdots & 0\\ \boldsymbol{CA}(k+1)\boldsymbol{B}(k) & \boldsymbol{CB}(k+1) & \cdots & 0\\ \vdots & \vdots & \ddots & \vdots\\ \boldsymbol{C}\prod_{i=1}^{N-1}\boldsymbol{A}(k+i)\boldsymbol{B}(k) & \boldsymbol{C}\prod_{i=2}^{N-1}\boldsymbol{A}(k+i)\boldsymbol{B}(k+1) & \cdots & \boldsymbol{CB}(k+N-1)\end{bmatrix}$$

$\tilde{\boldsymbol{\Phi}}(k)=\boldsymbol{\Lambda}_N\boldsymbol{\Phi}(k)=$

$$\begin{bmatrix}\boldsymbol{CB}(k) & 0 & \cdots & 0\\ \boldsymbol{CA}(k+1)\boldsymbol{B}(k)+\boldsymbol{CB}(k) & \boldsymbol{CB}(k+1) & \cdots & 0\\ \vdots & \vdots & \ddots & \vdots\\ \sum_{j=1}^{N-1}\boldsymbol{C}\prod_{i=1}^{j}\boldsymbol{A}(k+i)\boldsymbol{B}(k)+\boldsymbol{CB}(k) & \sum_{j=2}^{N-1}\boldsymbol{C}\prod_{i=2}^{j}\boldsymbol{A}(k+i)\boldsymbol{B}(k+1)+\boldsymbol{CB}(k+1) & \cdots & \boldsymbol{CB}(k+N-1)\end{bmatrix}$$

$$\boldsymbol{\Psi}(k)=\begin{bmatrix}\boldsymbol{CA}(k)\\ \boldsymbol{CA}(k+1)\boldsymbol{A}(k)\\ \vdots\\ \boldsymbol{C}\prod_{i=0}^{N-1}\boldsymbol{A}(k+i)\end{bmatrix},\quad \tilde{\boldsymbol{\Psi}}(k)=\boldsymbol{\Lambda}_N\boldsymbol{\Psi}(k)=\begin{bmatrix}\boldsymbol{CA}(k)\\ \boldsymbol{CA}(k+1)\boldsymbol{A}(k)+\boldsymbol{CA}(k)\\ \vdots\\ \sum_{j=0}^{N-1}\boldsymbol{C}\prod_{i=0}^{j}\boldsymbol{A}(k+i)\end{bmatrix},$$

$$\boldsymbol{Y}_N(k+1)=\begin{bmatrix}y(k+1)\\ \vdots\\ y(k+N)\end{bmatrix}_{N\times1},\quad \boldsymbol{E}=\begin{bmatrix}1\\ \vdots\\ 1\end{bmatrix}_{N\times1},\quad \boldsymbol{\Lambda}_N=\begin{bmatrix}1 & & \\ \vdots & \ddots & \\ 1 & \cdots & 1\end{bmatrix}_{N\times N},$$

$$\Delta\boldsymbol{U}_N(k)=\begin{bmatrix}\Delta u(k)\\ \vdots\\ \Delta u(k+N-1)\end{bmatrix}_{N\times1},\quad \Delta\boldsymbol{Y}_N(k+1)=\boldsymbol{Y}_N(k+1)-\boldsymbol{Y}_N(k)$$

We can rewrite prediction model (5) into (6).

$$\Delta\boldsymbol{Y}_N(k+1)=\boldsymbol{\Psi}(k)\Delta\boldsymbol{x}(k)+\boldsymbol{\Phi}(k)\Delta\boldsymbol{U}_N(k)\tag{6}$$

Both sides of equation (6) are left multiplied by $\boldsymbol{\Lambda}_N$ to have

$$\begin{aligned}\boldsymbol{Y}_N(k+1)&=\boldsymbol{E}y(k)+\boldsymbol{\Lambda}_N\boldsymbol{\Psi}(k)\Delta\boldsymbol{x}(k)+\boldsymbol{\Lambda}_N\boldsymbol{\Phi}(k)\Delta\boldsymbol{U}_N(k)\\ &=\boldsymbol{E}y(k)+\tilde{\boldsymbol{\Psi}}(k)\Delta\boldsymbol{x}(k)+\tilde{\boldsymbol{\Phi}}(k)\Delta\boldsymbol{U}_N(k)\end{aligned}\tag{7}$$

### C. *Design of predictive control and performance analysis for SISO systems*

A weighted cost function is shown as

$$\begin{aligned}J=&\left[\boldsymbol{Y}_N^*(k+1)-\boldsymbol{Y}_N(k+1)\right]^T\boldsymbol{Q}\left[\boldsymbol{Y}_N^*(k+1)-\boldsymbol{Y}_N(k+1)\right]\\ &+\Delta\boldsymbol{U}_N^T(k)\boldsymbol{\lambda}\Delta\boldsymbol{U}_N(k)\end{aligned}\tag{8}$$

where $\tilde{\boldsymbol{Y}}_N^*(k+1)=\left[y^*(k+1),\cdots,y^*(k+N)\right]^T$ is the desired system output signal vector and $y^*(k+i)$ is the desired system output at the time $(k+i)$ $(i=1,2,\cdots,N)$. $\boldsymbol{\lambda}$=diag($\lambda_1,\cdots,\lambda_N$) and $\boldsymbol{Q}$=diag($q_1,\cdots,q_N$) are the weighted diagonal matrices and we normally assume that $\lambda_i(i=1,\cdots,N)$ are equal to $\lambda$.

We may obtain the unconstrained implicit MPC (uiMPC) vector $\Delta\boldsymbol{U}_N(k)$ by minimizing the cost function (9) or obtain the constrained implicit MPC (ciMPC) vector by minimizing the cost function (10) which is subject to constraints $g_1(\boldsymbol{U}_N(k))\in\Omega_1$ and $g_2(\boldsymbol{Y}_N(k))\in\Omega_2$.

$$\min_{\Delta\boldsymbol{U}_N(k)} J\tag{9}$$

$$\min_{\Delta\boldsymbol{U}_N(k)\ \text{s.t.}\ \boldsymbol{g}_1(\boldsymbol{U}_N(k))\in\Omega_1,\boldsymbol{g}_2(\boldsymbol{Y}_N(k+1))\in\Omega_2} J\tag{10}$$

We may substitute (7) into the cost function (8) and solve the optimization condition $\partial J/\partial\Delta\boldsymbol{U}_N(k)=0$ to have the unconstrained explicit MPC (11) which is the optimal solution of (9).

$$\begin{aligned}\Delta\boldsymbol{U}_N(k)=&[\tilde{\boldsymbol{\Phi}}^T(k)\boldsymbol{Q}\tilde{\boldsymbol{\Phi}}(k)+\boldsymbol{\lambda}]^{-1}\tilde{\boldsymbol{\Phi}}^T(k)\boldsymbol{Q}[\boldsymbol{Y}_N^*(k+1)\\ &-\boldsymbol{E}y(k)-\tilde{\boldsymbol{\Psi}}(k)\Delta\boldsymbol{x}(k)]\end{aligned}\tag{11}$$

Then the current input is given by

$$u(k)=u(k-1)+\boldsymbol{g}^T\Delta\boldsymbol{U}_{Nu}(k)\tag{12}$$

where $\boldsymbol{g}=\left[1,0,\cdots,0\right]^T$.

*Remark 1:* $\tilde{\boldsymbol{\Psi}}(k)$, $\tilde{\boldsymbol{\Phi}}(k)$ and $\tilde{\boldsymbol{\Phi}}_w(k)$ in (7) contain the predictive PG vector $\boldsymbol{\phi}_{Ly}^T(k+i)$ and $\boldsymbol{\phi}_{Lu}^T(k+i)$ ($i$= 0, 1, 2, $\cdots$, $N-1$) which are determined during the optimization process for (9) or (10). For inexpensive computation, we can make an approximation $\boldsymbol{\phi}_L^T(k+i)=\boldsymbol{\phi}_L^T(k)$ to have $\boldsymbol{A}(k+i)$=$\boldsymbol{A}(k)$ and $\boldsymbol{B}(k+i)$=$\boldsymbol{B}(k)$ if the system model is not intensively nonlinear or especially linear.

On the other hand, if the system is intensively nonlinear, we also recommend an alternative method: iteration control applied in [21], [23], [24].

### D. *Performance Analysis*

This section provides two kinds of the performance analyses of the system controlled by unconstrained MPC.

We define

$$\boldsymbol{\phi}_{Ly}(z^{-1})=\phi_1(k)+\cdots+\phi_{Ly}(k)z^{-Ly+1}\tag{13}$$

$$\boldsymbol{\phi}_{Lu}(z^{-1})=\phi_{Ly+1}(k)+\cdots+\phi_{Ly+Lu}(k)z^{-Lu+1}\tag{14}$$

$$\boldsymbol{P}(k)=[\tilde{\boldsymbol{\Phi}}^T(k)\boldsymbol{Q}\tilde{\boldsymbol{\Phi}}(k)+\boldsymbol{\lambda}]^{-1}\tilde{\boldsymbol{\Phi}}^T(k)\boldsymbol{Q}\tag{15}$$

$$\left[\left[\tilde{\boldsymbol{\Psi}}_1(k)\right]_{N\times Ly}\quad\left[\tilde{\boldsymbol{\Psi}}_2(k)\right]_{N\times Lu}\right]=\tilde{\boldsymbol{\Psi}}(k)\tag{16}$$

where $z^{-1}$ is the backward shift operator which implies $\Delta=1-z^{-1}$.

Then (3) is rewritten as

$$\Delta\boldsymbol{y}(k+1)=\boldsymbol{\phi}_{Ly}(z^{-1})\Delta\boldsymbol{y}(k)+\boldsymbol{\phi}_{Lu}(z^{-1})\Delta\boldsymbol{u}(k)\tag{17}$$

**Performance Analysis 1**: Combining (11)-(17) yields the following transient closed-loop system equation concerning $y(k+1)$:

$$\begin{aligned}&\Big[(1-z^{-1}\boldsymbol{\phi}_{Ly}(z^{-1}))\Delta+z^{-1}\boldsymbol{\phi}_{Lu}(z^{-1})(1+z^{-1}\boldsymbol{P}(k)\tilde{\boldsymbol{\Psi}}_2(k)\boldsymbol{T}_u)^{-1}\boldsymbol{P}(k)\tilde{\boldsymbol{\Psi}}_1(k)\boldsymbol{T}_y\Delta\\ &\quad+z^{-1}\boldsymbol{\phi}_{Lu}(z^{-1})(1+z^{-1}\boldsymbol{P}(k)\tilde{\boldsymbol{\Psi}}_2(k)\boldsymbol{T}_u)^{-1}\boldsymbol{P}(k)\boldsymbol{E}\Big]y(k+1)\\ &=\boldsymbol{\phi}_{Lu}(z^{-1})(1+z^{-1}\boldsymbol{P}(k)\tilde{\boldsymbol{\Psi}}_2(k)\boldsymbol{T}_u)^{-1}\boldsymbol{P}(k)\boldsymbol{H}y^*(k+1)\end{aligned}\tag{18}$$

where $\boldsymbol{H}=[1,z,\cdots,z^{N-1}]^T$, $\boldsymbol{T}_y=[1,\cdots,z^{-Ly+1}]^T$ and $\boldsymbol{T}_u=[1,z^{-1},\cdots,z^{-Lu+1}]^T$.

We may choose the appropriate $N$, $\boldsymbol{Q}$ and $\boldsymbol{\lambda}$ so that inequality (19) holds

$$\begin{aligned}T(z^{-1})=&\Big[(1-z^{-1}\boldsymbol{\phi}_{Ly}(z^{-1}))\Delta\\ &+z^{-1}\boldsymbol{\phi}_{Lu}(z^{-1})(1+z^{-1}\boldsymbol{P}(k)\tilde{\boldsymbol{\Psi}}_2(k)\boldsymbol{T}_u)^{-1}\boldsymbol{P}(k)\tilde{\boldsymbol{\Psi}}_1(k)\boldsymbol{T}_y\Delta\\ &+z^{-1}\boldsymbol{\phi}_{Lu}(z^{-1})(1+z^{-1}\boldsymbol{P}(k)\tilde{\boldsymbol{\Psi}}_2(k)\boldsymbol{T}_u)^{-1}\boldsymbol{P}(k)\boldsymbol{E}\Big]\neq0\qquad|z|>1\end{aligned}\tag{19}$$

to guarantee the roots of the characteristic equation within the unit loop for the system stability.

We assume that the system is stable so that the steady-state error in response to certain desired system outputs $y^*(k+1)$ sometimes can be determined by

$$\begin{aligned}&\lim_{k\to\infty}e(k)\\ &=\lim_{z\to1}(1-z^{-1})T^{-1}(z^{-1})\Big[(1-z^{-1}\boldsymbol{\phi}_{Ly}(z^{-1}))\Delta\\ &\quad+z^{-1}\boldsymbol{\phi}_{Lu}(z^{-1})(1+z^{-1}\boldsymbol{P}(k)\tilde{\boldsymbol{\Psi}}_2(k)\boldsymbol{T}_u)^{-1}\boldsymbol{P}(k)\tilde{\boldsymbol{\Psi}}_1(k)\boldsymbol{T}_y\Delta\\ &\quad-\boldsymbol{\phi}_{Lu}(z^{-1})(1+z^{-1}\boldsymbol{P}(k)\tilde{\boldsymbol{\Psi}}_2(k)\boldsymbol{T}_u)^{-1}\boldsymbol{P}(k)(\boldsymbol{H}-z^{-1}\boldsymbol{E})\Big]\boldsymbol{Z}(y^*(k+1))\end{aligned}\tag{20}$$

where $\boldsymbol{Z}$ represents "the $z$ transformation of".

**Performance Analysis 2**: Combining (7) and (11) yields

$$\boldsymbol{\lambda}\Delta\boldsymbol{U}_N(k) = \tilde{\boldsymbol{\Phi}}^T(k)\boldsymbol{Q}[\boldsymbol{Y}_N^*(k+1) - \boldsymbol{Y}_N(k+1)] \tag{21}$$

Combining (12), (17) and (21) yields the following transient closed-loop system equation:

$$\begin{aligned}&\left[\lambda\Delta[I - z^{-1}\boldsymbol{\phi}_{Ly}(z^{-1})] + \boldsymbol{\phi}_{Lu}(z^{-1})\boldsymbol{g}^T\tilde{\boldsymbol{\Phi}}^T(k)\boldsymbol{QH}\right]y(k+1)\\ &= \boldsymbol{\phi}_{Lu}(z^{-1})\boldsymbol{g}^T\tilde{\boldsymbol{\Phi}}^T(k)\boldsymbol{QH}y^*(k+1)\end{aligned} \tag{22}$$

We may choose the appropriate $N$, $\boldsymbol{Q}$ and $\boldsymbol{\lambda}$ so that inequality (23) holds

$$T_1(z^{-1}) = \left[\lambda\Delta[I - z^{-1}{}_{Ly}(z^{-1})] + {}_{Lu}(z^{-1})\boldsymbol{g}^T\tilde{\boldsymbol{\Phi}}^T(k)\boldsymbol{QH}\right] \neq 0 \quad |z| > 1 \tag{23}$$

We assume that the system is stable so that the predictive error in response to certain desired system outputs $y^*(k+1)$ may be determined by

$$\begin{aligned}&\boldsymbol{\phi}_{Lu}(z^{-1})\boldsymbol{g}^T\tilde{\boldsymbol{\Phi}}^T(k)\boldsymbol{QH}e(k+1) + \lambda\Delta[I - z^{-1}\boldsymbol{\phi}_{Ly}(z^{-1})]e(k+1)\\ &= \lambda\Delta[I - z^{-1}\boldsymbol{\phi}_{Ly}(z^{-1})]y^*(k+1)\end{aligned} \tag{24}$$

From (24), we can conclude that the static error is eliminated when $\lambda=0$. On the other hand, we may not determine the static error through (24) when $\lambda\neq0$, since it contains the predictive output of system $y(k+i)$ in future time $k+i$ ($i=2,\cdots,N$) which does not happen. Additionally, if we choose the predictive step equal to time delay $N=d$, the controller (11) will degenerate into the incremental form of generalized minimum variance control. Only in this case, we may calculate the steady-state error in accordance with (24), owning to the absence of the predictive output of system $y(k+i)$, ($i=2,\cdots,N$) in (24).

The proposed MPC method and conclusions can be extended into MIMO systems naturally.

### E. Simulation

*Example 1*: One may design the explicit predictive controller (11), (12) for the following simple linear system model (25) to obtain the desired system performance characteristics according to (23), (24).

$$y(k) = 0.8y(k-2) + u(k-4) + 0.5u(k-5) \tag{25}$$

We choose the controller parameters with $N=4$, $\boldsymbol{Q}=\boldsymbol{I}$ (or $\boldsymbol{Q}$=[0 0 0 0; 0 0 0 0; 0 0 0 0; 0 0 0 1]) to have the steady-state error in response to the unit-ramp desired system output $y^*(k)$ as

$$\lim_{k\to\infty} e(k) = \frac{2\lambda}{15} \tag{26}$$

Table I lists the measured tracking errors of the system in simulations under the different values of $\lambda$, which firstly validates our conclusions for the SISO linear system.

TABLE I Tracking error $e(k)$ of the system under the different $\lambda$ values

| $\lambda$ | 0.1 | 1 | 2 |
|---|---|---|---|
| $e(200)=\cdots=e(700)$ | 0.013333333333 | 0.133333333333 | 0.266666666666 |

## III. System Description and Predictive Control Design for MIMO Systems

### A. System description for MIMO systems with disturbance

In this section, the EDLM with disturbance for MIMO systems is given as a fundamental tool for the predictive control design, and its basic assumptions and theorem are given.

We consider the following discrete-time MIMO process, and the relevant nonlinear system is shown as follows:

$$\boldsymbol{y}(k+1) = \boldsymbol{f}(\boldsymbol{y}(k),\cdots,\boldsymbol{y}(k-n_y),\boldsymbol{u}(k),\cdots,\boldsymbol{u}(k-n_u)) + \boldsymbol{w}(k+1) \tag{27}$$

where $\boldsymbol{f}(\cdots)=[f_1(\cdots),\cdots, f_{My}(\cdots)]^T\in\boldsymbol{R}^{My}$ is the nonlinear vector-valued differentiable function; $\boldsymbol{w}(k)$ denotes the disturbance vector. The dimensions of $\boldsymbol{y}(k)$ and $\boldsymbol{w}(k)$ are $M_y$, and $\boldsymbol{u}(k)$ is $M_u$.

Define $\boldsymbol{\varphi}(k) = [\boldsymbol{y}(k),\cdots,\boldsymbol{y}(k-n_y),\boldsymbol{u}(k),\cdots,\boldsymbol{u}(k-n_u)]$ to rewrite (27) as

$$\boldsymbol{y}(k+1) = \boldsymbol{f}(\boldsymbol{\varphi}(k)) + \boldsymbol{w}(k+1) \tag{28}$$

*Theorem 2:* If $\Delta\boldsymbol{H}(k)\neq\boldsymbol{0}$, there must exist a time-varying matrix $\boldsymbol{\phi}_L^T(k)$ called pseudo Jacobi matrix (PJM) and the system (27) can be transformed into the EDLM with disturbance shown as follow:

$$\Delta\boldsymbol{y}(k+1) = \boldsymbol{\phi}_L^T(k)\Delta\boldsymbol{H}(k) + \Delta\boldsymbol{w}(k+1) \tag{29}$$

where

$\Delta\boldsymbol{H}(k) = \left[\Delta\boldsymbol{Y}_{Ly}^T(k) \quad \Delta\boldsymbol{U}_{Lu}^T(k)\right]^T$;

$\Delta\boldsymbol{Y}_{Ly}(k) = [\Delta\boldsymbol{y}^T(k),\cdots,\Delta\boldsymbol{y}^T(k-L_y+1)]^T$;

$\Delta\boldsymbol{U}_{Lu}(k) = [\Delta\boldsymbol{u}^T(k),\cdots,\Delta\boldsymbol{u}^T(k-L_u+1)]^T$;

$\Delta\boldsymbol{w}(k+1) = [\Delta w_1(k+1),\cdots,\Delta w_{My}(k+N)]^T$

$\boldsymbol{\phi}_L^T(k) = [\boldsymbol{\phi}_{Ly}^T(k),\boldsymbol{\phi}_{Lu}^T]$, $\boldsymbol{\phi}_{Ly}^T(k) = [\boldsymbol{\Phi}_1(k),\cdots,\boldsymbol{\Phi}_{Ly}(k)]_{My\times(Ly\bullet My)}$,

$\boldsymbol{\phi}_{Lu}^T(k) = [\boldsymbol{\Phi}_{Ly+1}(k),\cdots,\boldsymbol{\Phi}_{Ly+Lu}(k)]_{My\times(Lu\bullet Mu)}$,

$\boldsymbol{\Phi}_i(k)\in\boldsymbol{R}^{My\times My}$ ($i=1,\cdots,L_y$); $\boldsymbol{\Phi}_i(k)\in\boldsymbol{R}^{My\times Mu}$ ($i=L_y+1,\cdots,L_y+L_u$);

In this paper, we only analyze the case $L_y=n_y+1$ and $L_u=n_u+1$.

*Proof:* Similar to Theorem 1.

### B. Predictive model

We can rewrite (29) into (30).

$$\begin{aligned}\Delta\boldsymbol{x}(k+1) &= \boldsymbol{A}(k)\Delta\boldsymbol{x}(k) + \boldsymbol{B}(k)\Delta\boldsymbol{u}(k) + \boldsymbol{T}\Delta\boldsymbol{w}(k+1)\\ \Delta\boldsymbol{y}(k+1) &= \boldsymbol{C}\Delta\boldsymbol{x}(k+1)\end{aligned} \tag{30}$$

where

$\boldsymbol{B}_{(Ly\bullet My+Lu\bullet Mu)\times Mu} = \left[\boldsymbol{\Phi}_{Ly+1}^T(k),\boldsymbol{0},\cdots,\boldsymbol{0},\boldsymbol{I},\boldsymbol{0},\cdots,\boldsymbol{0}\right]^T$,

$\boldsymbol{C} = \boldsymbol{T}^T = [\boldsymbol{I},\boldsymbol{0},\cdots,\boldsymbol{0}]_{My\times(Ly\bullet My+Lu\bullet Mu)}$, $\Delta\boldsymbol{x}(k) = \begin{bmatrix}\Delta\boldsymbol{Y}_{Ly}(k)\\ \Delta\boldsymbol{U}_{Lu}(k-1)\end{bmatrix}$,

$\boldsymbol{A}(k) =$

$$\begin{bmatrix}
\boldsymbol{\Phi}_1(k) & \boldsymbol{\Phi}_2(k) & \cdots & \boldsymbol{\Phi}_{Ly-1}(k) & \boldsymbol{\Phi}_{Ly}(k) & \boldsymbol{\Phi}_{Ly+2}(k) & \cdots & \boldsymbol{\Phi}_{Ly+Lu}(k) & \boldsymbol{0}\\
\boldsymbol{I}_{My\times My} & \boldsymbol{0} & \cdots & \boldsymbol{0} & \boldsymbol{0} & \boldsymbol{0} & \cdots & \boldsymbol{0} & \boldsymbol{0}\\
\boldsymbol{0} & \boldsymbol{I}_{My\times My} & \cdots & \boldsymbol{0} & \boldsymbol{0} & \boldsymbol{0} & \ddots & \boldsymbol{0} & \boldsymbol{0}\\
\vdots & \vdots & \ddots & \vdots & \vdots & \vdots & \ddots & \vdots & \vdots\\
\boldsymbol{0} & \boldsymbol{0} & \cdots & \boldsymbol{I}_{My\times My} & \boldsymbol{0} & \boldsymbol{0} & \cdots & \boldsymbol{0} & \boldsymbol{0}\\
\boldsymbol{0} & \boldsymbol{0} & \cdots & \boldsymbol{0} & \boldsymbol{0} & \boldsymbol{0} & \cdots & \boldsymbol{0} & \boldsymbol{0}\\
\boldsymbol{0} & \boldsymbol{0} & \cdots & \boldsymbol{0} & \boldsymbol{0} & \boldsymbol{I}_{Mu\times Mu} & \ddots & \boldsymbol{0} & \boldsymbol{0}\\
\vdots & \vdots & \ddots & \vdots & \vdots & \vdots & \ddots & \vdots & \vdots\\
\boldsymbol{0} & \boldsymbol{0} & \cdots & \boldsymbol{0} & \boldsymbol{0} & \boldsymbol{0} & \cdots & \boldsymbol{I}_{Mu\times Mu} & \boldsymbol{0}
\end{bmatrix}$$

Then we have finite $N$ step forward prediction model as follow:

$$
\begin{aligned}
\Delta\boldsymbol{x}(k+1) &= \boldsymbol{A}(k)\Delta\boldsymbol{x}(k)+\boldsymbol{B}(k)\Delta\boldsymbol{u}(k)+\boldsymbol{T}\Delta\boldsymbol{w}(k+1)\\
\Delta\boldsymbol{x}(k+2) &= \boldsymbol{A}(k+1)\Delta\boldsymbol{x}(k+1)+\boldsymbol{B}(k+1)\Delta\boldsymbol{u}(k+1)+\boldsymbol{T}\Delta\boldsymbol{w}(k+2)\\
&= \boldsymbol{A}(k+1)\boldsymbol{A}(k)\Delta\boldsymbol{x}(k)+\boldsymbol{A}(k+1)\boldsymbol{B}(k)\Delta\boldsymbol{u}(k)\\
&\quad +\boldsymbol{B}(k+1)\Delta\boldsymbol{u}(k+1)+\boldsymbol{A}(k+1)\boldsymbol{T}\Delta\boldsymbol{w}(k+1)\\
&\quad +\boldsymbol{T}\Delta\boldsymbol{w}(k+2)\\
&\ \ \vdots\\
\Delta\boldsymbol{x}(k+N) &= \boldsymbol{A}(k+N-1)\Delta\boldsymbol{x}(k+N-1)\\
&\quad +\boldsymbol{B}(k+N-1)\Delta\boldsymbol{u}(k+N-1)\ +\boldsymbol{T}\Delta\boldsymbol{w}(k+N)\\
&= \prod_{i=0}^{N-1}\boldsymbol{A}(k+i)\Delta\boldsymbol{x}(k)+\prod_{i=1}^{N-1}\boldsymbol{A}(k+i)\boldsymbol{B}(k)\Delta\boldsymbol{u}(k)\\
&\quad +\prod_{i=2}^{N-1}\boldsymbol{A}(k+i)\boldsymbol{B}(k+1)\Delta\boldsymbol{u}(k+1)+\cdots\\
&\quad +\boldsymbol{B}(k+N-1)\Delta\boldsymbol{u}(k+N-1)\\
&\quad +\prod_{i=1}^{N-1}\boldsymbol{A}(k+i)\bullet\boldsymbol{T}\Delta\boldsymbol{w}(k+1)\\
&\quad +\prod_{i=2}^{N-1}\boldsymbol{A}(k+i)\bullet\boldsymbol{T}\Delta\boldsymbol{w}(k+2)+\cdots+\boldsymbol{T}\Delta\boldsymbol{w}(k+N)
\end{aligned}
\tag{31}
$$

Here, we define $\boldsymbol{Y}_N(k)$, $\Delta\boldsymbol{Y}_N(k+1)$, $\Delta\boldsymbol{U}_N(k)$, $\boldsymbol{\Psi}(k)$, $\tilde{\boldsymbol{\Psi}}(k)$, $\boldsymbol{\Phi}(k)$, $\tilde{\boldsymbol{\Phi}}(k)$ as follows:

$$
\boldsymbol{\Phi}(k)=
\begin{bmatrix}
\boldsymbol{CB}(k) & \boldsymbol{0} & \cdots & \boldsymbol{0}\\
\boldsymbol{CA}(k+1)\boldsymbol{B}(k) & \boldsymbol{CB}(k+1) & \cdots & \boldsymbol{0}\\
\vdots & \vdots & \ddots & \vdots\\
\boldsymbol{C}\prod_{i=1}^{N-1}\boldsymbol{A}(k+i)\boldsymbol{B}(k) & \boldsymbol{C}\prod_{i=2}^{N-1}\boldsymbol{A}(k+i)\boldsymbol{B}(k+1) & \cdots & \boldsymbol{CB}(k+N-1)
\end{bmatrix}
$$

$$
\tilde{\boldsymbol{\Phi}}(k)=\boldsymbol{\Lambda}_N\boldsymbol{\Phi}(k)=
\begin{bmatrix}
\boldsymbol{CB}(k) & \boldsymbol{0} & \cdots & \boldsymbol{0}\\
\boldsymbol{CA}(k+1)\boldsymbol{B}(k)+\boldsymbol{CB}(k) & \boldsymbol{CB}(k+1) & \cdots & \boldsymbol{0}\\
\vdots & \vdots & \ddots & \vdots\\
\sum_{j=1}^{N-1}\boldsymbol{C}\prod_{i=1}^{j}\boldsymbol{A}(k+i)\boldsymbol{B}(k)+\boldsymbol{CB}(k) & \sum_{j=2}^{N-1}\boldsymbol{C}\prod_{i=2}^{j}\boldsymbol{A}(k+i)\boldsymbol{B}(k+1)+\boldsymbol{CB}(k+1) & \cdots & \boldsymbol{CB}(k+N-1)
\end{bmatrix}
$$

$$
\boldsymbol{\Psi}(k)=
\begin{bmatrix}
\boldsymbol{CA}(k)\\
\boldsymbol{CA}(k+1)\boldsymbol{A}(k)\\
\vdots\\
\boldsymbol{C}\prod_{i=0}^{N-1}\boldsymbol{A}(k+i)
\end{bmatrix},\quad
\tilde{\boldsymbol{\Psi}}(k)=\boldsymbol{\Lambda}_N\tilde{\boldsymbol{\Psi}}(k)=
\begin{bmatrix}
\boldsymbol{CA}(k)\\
\boldsymbol{CA}(k+1)\boldsymbol{A}(k)+\boldsymbol{CA}(k)\\
\vdots\\
\sum_{j=0}^{N-1}\boldsymbol{C}\prod_{i=0}^{j}\boldsymbol{A}(k+i)
\end{bmatrix},
$$

$$
\boldsymbol{\Phi}_w(k)=
\begin{bmatrix}
\boldsymbol{CT} & \boldsymbol{0} & \cdots & \boldsymbol{0}\\
\boldsymbol{CA}(k+1)\boldsymbol{T} & \boldsymbol{CT} & \cdots & \boldsymbol{0}\\
\vdots & \vdots & \ddots & \vdots\\
\boldsymbol{C}\prod_{i=1}^{N-1}\boldsymbol{A}(k+i)\boldsymbol{T} & \boldsymbol{C}\prod_{i=2}^{N-1}\boldsymbol{A}(k+i)\boldsymbol{T} & \cdots & \boldsymbol{CT}
\end{bmatrix},
$$

$$
\tilde{\boldsymbol{\Phi}}_w(k)=\boldsymbol{\Lambda}_N\boldsymbol{\Phi}_w(k)
=\begin{bmatrix}
\boldsymbol{CT} & \boldsymbol{0} & \cdots & \boldsymbol{0}\\
\boldsymbol{CA}(k+1)\boldsymbol{T}+\boldsymbol{CT} & \boldsymbol{CT} & \cdots & \boldsymbol{0}\\
\vdots & \vdots & \ddots & \vdots\\
\sum_{j=1}^{N-1}\boldsymbol{C}\prod_{i=1}^{j}\boldsymbol{A}(k+i)\boldsymbol{T}+\boldsymbol{CT} & \sum_{j=2}^{N-1}\boldsymbol{C}\prod_{i=2}^{j}\boldsymbol{A}(k+i)\boldsymbol{T}+\boldsymbol{CT} & \cdots & \boldsymbol{CT}
\end{bmatrix},
$$

$$
\boldsymbol{Y}_N(k+1)=
\begin{bmatrix}
\boldsymbol{y}(k+1)\\ \vdots\\ \boldsymbol{y}(k+N)
\end{bmatrix}_{(N\bullet My)\times 1},\quad
\boldsymbol{\Lambda}_N=
\begin{bmatrix}
\boldsymbol{I} & & \\ \vdots & \ddots & \\ \boldsymbol{I} & \cdots & \boldsymbol{I}
\end{bmatrix}_{(N\bullet My)\times(N\bullet My)},
$$

$$
\Delta\boldsymbol{U}_N(k)=
\begin{bmatrix}
\Delta\boldsymbol{u}(k)\\ \vdots\\ \Delta\boldsymbol{u}(k+N-1)
\end{bmatrix}_{(N\bullet Mu)\times 1},\quad
\Delta\boldsymbol{W}(k+1)=
\begin{bmatrix}
\Delta\boldsymbol{w}(k+1)\\ \vdots\\ \Delta\boldsymbol{w}(k+N)
\end{bmatrix}.
$$

$$
\boldsymbol{E}=
\begin{bmatrix}
\boldsymbol{I}\\ \vdots\\ \boldsymbol{I}
\end{bmatrix}_{(N\bullet My)\times My},\quad
\Delta\boldsymbol{Y}_N(k+1)=\boldsymbol{Y}_N(k+1)-\boldsymbol{Y}_N(k).
$$

Then the prediction model (31) can be written as (32):

$$
\Delta\boldsymbol{Y}_N(k+1)=\boldsymbol{\Psi}(k)\Delta\boldsymbol{x}(k)+\boldsymbol{\Phi}(k)\Delta\boldsymbol{U}_N(k)+\boldsymbol{\Phi}_w(k)\Delta\boldsymbol{W}(k+1)
\tag{32}
$$

Both sides of equation (32) are left multiplied by $\boldsymbol{\Lambda}_N$ to have

$$
\begin{aligned}
\boldsymbol{Y}_N(k+1)&=\boldsymbol{Ey}(k)+\tilde{\boldsymbol{\Psi}}(k)\Delta\boldsymbol{x}(k)+\tilde{\boldsymbol{\Phi}}(k)\Delta\boldsymbol{U}_N(k)\\
&\quad+\tilde{\boldsymbol{\Phi}}_w(k)\Delta\boldsymbol{W}(k+1)
\end{aligned}
\tag{33}
$$

### *C. Design of predictive control compensated with disturbance and performance analysis for MIMO systems*

A weighted cost function is shown as

$$
\begin{aligned}
J&=\left[\boldsymbol{Y}_N^*(k+1)-\boldsymbol{Y}_N(k+1)\right]^T\boldsymbol{Q}\left[\boldsymbol{Y}_N^*(k+1)-\boldsymbol{Y}_N(k+1)\right]\\
&\quad+\Delta\boldsymbol{U}_N^T(k)\boldsymbol{\lambda}\Delta\boldsymbol{U}_N(k)
\end{aligned}
\tag{34}
$$

where $\tilde{\boldsymbol{Y}}_N^*(k+1)=\left[\boldsymbol{y}^*(k+1),\cdots,\boldsymbol{y}^*(k+N)\right]^T$ is the desired system output signal vector. $\boldsymbol{\lambda}$=diag($\lambda_1$,⋯,$\lambda_{N\times Mu}$) and $\boldsymbol{Q}$=diag($q_1$, ⋯, $q_{N\times My}$) are the weighted diagonal matrices and we assume that $\lambda_i$($i$=1,⋯, $N\times M_u$) are equal to $\lambda$;

We may obtain the unconstrained implicit MPC compensated with disturbance (uiMPC+D) vector by minimizing the cost function (35) or obtain the constrained implicit MPC compensated with disturbance (ciMPC+D) vector by minimizing the cost function (36) which is subject to constraints $\boldsymbol{g}_1(\boldsymbol{U}_N(k))\in\Omega_1$ and $\boldsymbol{g}_2(\boldsymbol{Y}_N(k))\in\Omega_2$.

$$
\min_{\Delta\boldsymbol{U}_N(k)} J
\tag{35}
$$

$$
\min_{\Delta\boldsymbol{U}_N(k)\ \text{s.t.}\ \boldsymbol{g}_1(\boldsymbol{U}_N(k))\in\Omega_1,\ \boldsymbol{g}_2(\boldsymbol{Y}_N(k+1))\in\Omega_2} J
\tag{36}
$$

We may substitute (33) into the cost function (34) and solve the optimization condition $\partial J/\partial\Delta\boldsymbol{U}_N(k)=\boldsymbol{0}$ to have the unconstrained explicit predictive control compensated with disturbance (37).

$$
\begin{aligned}
\Delta\boldsymbol{U}_N(k)&=[\tilde{\boldsymbol{\Phi}}^T(k)\boldsymbol{Q}\tilde{\boldsymbol{\Phi}}(k)+\boldsymbol{\lambda}]^{-1}\tilde{\boldsymbol{\Phi}}^T(k)\boldsymbol{Q}[\boldsymbol{Y}_N^*(k+1)\\
&\quad-\boldsymbol{Ey}(k)-\tilde{\boldsymbol{\Psi}}(k)\Delta\boldsymbol{x}(k)-\tilde{\boldsymbol{\Phi}}_w(k)\Delta\boldsymbol{W}(k+1)]
\end{aligned}
\tag{37}
$$

Since the accurate disturbance vector may not be obtained directly, we rewrite controller (37) into (38) by defining $\hat{w}(k+1)$ as the measurement or estimation of $w(k+1)$ and $\hat{\boldsymbol{W}}(k+1)$ as that of $\boldsymbol{W}(k+1)$.

$$\begin{aligned}\Delta\boldsymbol{U}_N(k) = [\tilde{\boldsymbol{\Phi}}^T(k)\boldsymbol{Q}\tilde{\boldsymbol{\Phi}}(k)+\boldsymbol{\lambda}]^{-1}\tilde{\boldsymbol{\Phi}}^T(k)\boldsymbol{Q}[\boldsymbol{Y}_N^*(k+1)\\ -\boldsymbol{E}\boldsymbol{y}(k)-\tilde{\boldsymbol{\Psi}}(k)\Delta\boldsymbol{x}(k)-\tilde{\boldsymbol{\Phi}}_w(k)\Delta\hat{\boldsymbol{W}}(k+1)]\end{aligned} \tag{38}$$

Then the current input is given by

$$\boldsymbol{u}(k)=\boldsymbol{u}(k-1)+\boldsymbol{g}^T\Delta\boldsymbol{U}_{Nu}(k) \tag{39}$$

where $\boldsymbol{g}=\left[\boldsymbol{I},\boldsymbol{0},\cdots,\boldsymbol{0}\right]^T$.

*D. Performance Analysis*

This section provides two kinds of the performance analyses of the system controlled by unconstrained MPC.

Define

$$\boldsymbol{\phi}_{Ly}(z^{-1})=\boldsymbol{\Phi}_1(k)+\cdots+\boldsymbol{\Phi}_{Ly}(k)z^{-Ly+1} \tag{40}$$

$$\boldsymbol{\phi}_{Lu}(z^{-1})=\boldsymbol{\Phi}_{Ly+1}(k)+\cdots+\boldsymbol{\Phi}_{Ly+Lu}(k)z^{-Lu+1} \tag{41}$$

to rewrite (29) into

$$\Delta\boldsymbol{y}(k+1)=\boldsymbol{\phi}_{Ly}(z^{-1})\Delta\boldsymbol{y}(k)+\boldsymbol{\phi}_{Lu}(z^{-1})\Delta\boldsymbol{u}(k)+\Delta\boldsymbol{w}(k+1) \tag{42}$$

and define

$$\left[\left[\tilde{\boldsymbol{\Psi}}_1(k)\right]_{\substack{N\bullet My\\ \times Ly\bullet My}} \quad \left[\tilde{\boldsymbol{\Psi}}_2(k)\right]_{\substack{N\bullet My\\ \times Lu\bullet Mu}}\right]=\tilde{\boldsymbol{\Psi}}(k) \tag{43}$$

$$\boldsymbol{P}(k)=[\tilde{\boldsymbol{\Phi}}^T(k)\boldsymbol{Q}\tilde{\boldsymbol{\Phi}}(k)+\boldsymbol{\lambda}]^{-1}\tilde{\boldsymbol{\Phi}}^T(k)\boldsymbol{Q} \tag{44}$$

**Performance Analysis 1**: Combining (38)-(44) yields the following transient closed-loop system equation concerning $\boldsymbol{y}(k+1)$:

$$\begin{aligned}&\Big[(\boldsymbol{I}-z^{-1}\boldsymbol{\phi}_{Ly}(z^{-1}))\Delta+z^{-1}\boldsymbol{\phi}_{Lu}(z^{-1})(\boldsymbol{I}+z^{-1}\boldsymbol{P}(k)\tilde{\boldsymbol{\Psi}}_2(k)\boldsymbol{T}_u)^{-1}\boldsymbol{P}(k)\\ &\quad\bullet\tilde{\boldsymbol{\Psi}}_1(k)\boldsymbol{T}_y\Delta+z^{-1}\boldsymbol{\phi}_{Lu}(z^{-1})(\boldsymbol{I}+z^{-1}\boldsymbol{P}(k)\tilde{\boldsymbol{\Psi}}_2(k)\boldsymbol{T}_u)^{-1}\boldsymbol{P}(k)\boldsymbol{E}\Big]\boldsymbol{y}(k+1)\\ &=\boldsymbol{\phi}_{Lu}(z^{-1})(\boldsymbol{I}+z^{-1}\boldsymbol{P}(k)\tilde{\boldsymbol{\Psi}}_2(k)\boldsymbol{T}_u)^{-1}\boldsymbol{P}(k)\boldsymbol{H}\boldsymbol{y}^*(k+1)+\Delta\boldsymbol{w}(k+1)\\ &\quad-\boldsymbol{\phi}_{Lu}(z^{-1})(\boldsymbol{I}+z^{-1}\boldsymbol{P}(k)\boldsymbol{\Psi}_2(k)\boldsymbol{T}_u)^{-1}\boldsymbol{P}(k)\tilde{\boldsymbol{\Phi}}_w(k)\Delta\hat{\boldsymbol{W}}(k+1)\end{aligned} \tag{45}$$

where $\boldsymbol{H}=[\boldsymbol{I},z\boldsymbol{I},\cdots,z^{N-1}\boldsymbol{I}]^T$, $\boldsymbol{T}_y=[\boldsymbol{I},\cdots,z^{-Ly+1}\boldsymbol{I}]^T$ and $\boldsymbol{T}_u=[\boldsymbol{I},z^{-1}\boldsymbol{I},\cdots,z^{-Lu+1}\boldsymbol{I}]^T$.

We may choose the appropriate $N$, $\boldsymbol{Q}$ and $\boldsymbol{\lambda}$ so that inequality (46) holds

$$\begin{aligned}\boldsymbol{T}(z^{-1})&=\Big[(\boldsymbol{I}-z^{-1}\boldsymbol{\phi}_{Ly}(z^{-1}))\Delta\\ &\quad+z^{-1}\boldsymbol{\phi}_{Lu}(z^{-1})(\boldsymbol{I}+z^{-1}\boldsymbol{P}(k)\tilde{\boldsymbol{\Psi}}_2(k)\boldsymbol{T}_u)^{-1}\boldsymbol{P}(k)\tilde{\boldsymbol{\Psi}}_1(k)\boldsymbol{T}_y\Delta\\ &\quad+z^{-1}\boldsymbol{\phi}_{Lu}(z^{-1})(\boldsymbol{I}+z^{-1}\boldsymbol{P}(k)\tilde{\boldsymbol{\Psi}}_2(k)\boldsymbol{T}_u)^{-1}\boldsymbol{P}(k)\boldsymbol{E}\Big]\neq\boldsymbol{0}\qquad |z|>1\end{aligned} \tag{46}$$

to guarantee the roots of the characteristic equation within the unit loop for the system stability.

We assume $\Delta\hat{\boldsymbol{W}}(k+1)=\Delta\boldsymbol{W}(k+1)$ and the system stability is guaranteed, then the transient closed-loop pulse transfer function for the disturbance $\boldsymbol{w}(k+1)$ will be

$$\begin{aligned}&\boldsymbol{G}_w(z^{-1})\\ &=\frac{(1-z^{-1})(\boldsymbol{I}-\boldsymbol{\phi}_{Lu}(z^{-1})(\boldsymbol{I}+z^{-1}\boldsymbol{P}(k)\boldsymbol{\Psi}_2(k)\boldsymbol{T}_u)^{-1}\boldsymbol{P}(k)\tilde{\boldsymbol{\Phi}}_w(k)\boldsymbol{H})}{\boldsymbol{T}(z^{-1})}\end{aligned} \tag{47}$$

Further, the transient transfer function for disturbance will approach **0** as $\boldsymbol{\lambda}$ approaches **0**, and the influence of disturbance $\boldsymbol{w}(k)$ is theoretically removed by implicit MPC designed by minimizing (35) when $\boldsymbol{\lambda}=\boldsymbol{0}$.

On the other hand, if $\boldsymbol{w}(k+1)$ is unknown, we normally let $\Delta\hat{\boldsymbol{W}}(k+1)=\Delta\boldsymbol{W}(k+1)=\boldsymbol{0}$ in (38) and the transient closed-loop system equation will be

$$\begin{aligned}&\Big[(\boldsymbol{I}-z^{-1}\boldsymbol{\phi}_{Ly}(z^{-1}))\Delta+z^{-1}\boldsymbol{\phi}_{Lu}(z^{-1})(\boldsymbol{I}+z^{-1}\boldsymbol{P}(k)\tilde{\boldsymbol{\Psi}}_2(k)\boldsymbol{T}_u)^{-1}\boldsymbol{P}(k)\tilde{\boldsymbol{\Psi}}_1(k)\boldsymbol{T}_y\Delta\\ &\quad+z^{-1}\boldsymbol{\phi}_{Lu}(z^{-1})(\boldsymbol{I}+z^{-1}\boldsymbol{P}(k)\tilde{\boldsymbol{\Psi}}_2(k)\boldsymbol{T}_u)^{-1}\boldsymbol{P}(k)\boldsymbol{E}\Big]\boldsymbol{y}(k+1)\\ &=\boldsymbol{\phi}_{Lu}(z^{-1})(\boldsymbol{I}+z^{-1}\boldsymbol{P}(k)\tilde{\boldsymbol{\Psi}}_2(k)\boldsymbol{T}_u)^{-1}\boldsymbol{P}(k)\boldsymbol{H}\boldsymbol{y}^*(k+1)+\Delta\boldsymbol{w}(k+1)\end{aligned} \tag{48}$$

Furthermore, if rank$[\boldsymbol{\Phi}_{Ly+1}(k)]=M_y$ ($M_u\geq M_y$) and we choose $\boldsymbol{\lambda}=\boldsymbol{0}$, the closed-loop pulse transfer function for the disturbance $\boldsymbol{w}(k+1)$ will be

$$\boldsymbol{G}_w(z^{-1})=(1-z^{-1})\boldsymbol{I} \tag{49}$$

**Performance Analysis 2**: This is for the case when we choose the predictive step equal to time delay $N=d$. Combining (33) and (38) together yields

$$\begin{aligned}\boldsymbol{\lambda}\Delta\boldsymbol{U}_N(k)=\tilde{\boldsymbol{\Phi}}^T(k)\boldsymbol{Q}[\boldsymbol{Y}_N^*(k+1)-\boldsymbol{Y}_N(k+1)\\ +\tilde{\boldsymbol{\Phi}}_w(k)\Delta\boldsymbol{W}(k+1)-\tilde{\boldsymbol{\Phi}}_w(k)\Delta\hat{\boldsymbol{W}}(k+1)]\end{aligned} \tag{50}$$

Combining (39), (42) and (50) yields the following transient closed-loop system equation:

$$\begin{aligned}&\Big[\lambda\Delta[I-z^{-1}\boldsymbol{\phi}_{Ly}(z^{-1})]+\boldsymbol{\phi}_{Lu}(z^{-1})\boldsymbol{g}^T\tilde{\boldsymbol{\Phi}}^T(k)\boldsymbol{Q}\boldsymbol{H}\Big]\boldsymbol{y}(k+1)\\ &=\boldsymbol{\phi}_{Lu}(z^{-1})\boldsymbol{g}^T\tilde{\boldsymbol{\Phi}}^T(k)\boldsymbol{Q}\boldsymbol{H}\boldsymbol{y}^*(k+1)+\lambda\Delta\boldsymbol{w}(k+1)\\ &\quad+\boldsymbol{\phi}_{Lu}(z^{-1})\boldsymbol{g}^T\tilde{\boldsymbol{\Phi}}^T(k)\boldsymbol{Q}[\tilde{\boldsymbol{\Phi}}_w(k)\boldsymbol{H}\Delta\boldsymbol{w}(k+1)-\tilde{\boldsymbol{\Phi}}_w(k)\boldsymbol{H}\Delta\hat{\boldsymbol{w}}(k+1)]\end{aligned} \tag{51}$$

We may choose the appropriate $\boldsymbol{Q}$ and $\boldsymbol{\lambda}$ so that inequality (52) holds

$$\boldsymbol{T}_1(z^{-1})=\Big[\lambda\Delta[I-z^{-1}\boldsymbol{\phi}_{Ly}(z^{-1})]+\boldsymbol{\phi}_{Lu}(z^{-1})\boldsymbol{g}^T\tilde{\boldsymbol{\Phi}}^T(k)\boldsymbol{Q}\boldsymbol{H}\Big]\neq\boldsymbol{0}\quad |z|>1 \tag{52}$$

for the system stability.

We assume $\Delta\hat{\boldsymbol{W}}(k+1)=\Delta\boldsymbol{W}(k+1)$ and the system is stable, then the closed-loop pulse transfer function for the disturbance at time $k+1$ will be

$$\boldsymbol{G}(z^{-1})=\frac{\lambda(1-z^{-1})}{\lambda(1-z^{-1})[I-z^{-1}\boldsymbol{\phi}_{Ly}(z^{-1})]+\boldsymbol{\phi}_{Lu}(z^{-1})\boldsymbol{g}^T\tilde{\boldsymbol{\Phi}}^T(k)\boldsymbol{Q}\boldsymbol{H}} \tag{53}$$

Further, when we choose $\boldsymbol{\lambda}=\boldsymbol{0}$**,** the transient the closed-loop system equation will be

$$\begin{aligned}&\Big[\lambda\Delta[I-z^{-1}\boldsymbol{\phi}_{Ly}(z^{-1})]+\boldsymbol{\phi}_{Lu}(z^{-1})\boldsymbol{g}^T\tilde{\boldsymbol{\Phi}}^T(k)\boldsymbol{Q}\boldsymbol{H}\Big]\boldsymbol{y}(k+1)\\ &=\boldsymbol{\phi}_{Lu}(z^{-1})\boldsymbol{g}^T\tilde{\boldsymbol{\Phi}}^T(k)\boldsymbol{Q}\boldsymbol{H}\boldsymbol{y}^*(k+1)\end{aligned} \tag{54}$$

which means the influence of disturbance $\boldsymbol{w}(k)$ is theoretically eliminated when the system is stable.

On the other hand, if $\boldsymbol{w}(k+1)$ is unknown, we normally let $\Delta\hat{\boldsymbol{W}}(k+1)=\boldsymbol{0}$ in (38) and the transient closed-loop system equation will be

$$
\begin{aligned}
&\left[\lambda\Delta[\boldsymbol{I}-z^{-1}\boldsymbol{\phi}_{Ly}(z^{-1})]+\boldsymbol{\phi}_{Lu}(z^{-1})\boldsymbol{g}^T\tilde{\boldsymbol{\Phi}}^T(k)\boldsymbol{QH}\right]\boldsymbol{y}(k+1)\\
&=\boldsymbol{\phi}_{Lu}(z^{-1})\boldsymbol{g}^T\tilde{\boldsymbol{\Phi}}^T(k)\boldsymbol{QHy}^*(k+1)+\lambda\Delta\boldsymbol{w}(k+1)\\
&\quad+\boldsymbol{\phi}_{Lu}(z^{-1})\boldsymbol{g}^T\tilde{\boldsymbol{\Phi}}^T(k)\boldsymbol{Q}[\tilde{\boldsymbol{\Phi}}_w(k)\boldsymbol{H}\Delta\boldsymbol{w}(k+1)]
\end{aligned}
\tag{55}
$$

Furthermore, if rank[$\boldsymbol{\Phi}_{Ly+1}(k)$]=$M_y$ ($M_u \geq M_y$) and we choose $N$=1, the predictive control will degenerate into the controller in [25].

*E. Simulations*

*Example 2*: In this example, we want to show how to apply the proposed predictive controller in nonlinear systems. The system model is given as:

$$
\begin{aligned}
y_1(k) &= y_1^2(k-2)+0.7y_2^2(k-2)+u_1(k-3)+0.5u_2(k-3)\\
&\quad+0.4u_1^3(k-4)+0.5u_2(k-4)+w_1(k)\\
y_2(k) &= 0.5y_1^2(k-2)+1.3y_2^2(k-2)+0.4u_1(k-3)\\
&\quad+1.2u_2(k-3)+0.2u_1(k-4)+0.4u_2^3(k-4)+w_2(k)
\end{aligned}
\tag{56}
$$

The desired system output is

$$
\begin{aligned}
y_1^*(k) &= 0.2\sin(k/20)-0.2\sin(k/10)-0.2\cos(k/5)\\
&\quad+0.2\cos(k/2) \qquad 1\leq k\leq 350\\
y_2^*(k) &= -0.2\cos(k/15)-0.2\sin(k/25)+0.2\sin(k/5)\\
&\quad+0.2\cos(k/3) \qquad 1\leq k\leq 350\\
y_1^*(k) &= 0.5\times(-1)^{round(k+1/50)} \qquad 351\leq k\leq 700\\
y_2^*(k) &= -0.5\times(-1)^{round(k+1/50)} \qquad 351\leq k\leq 700
\end{aligned}
\tag{57}
$$

At beginning, we assume the disturbance vector $\boldsymbol{w}(k)=\boldsymbol{0}$. The initial values are $\boldsymbol{y}(1)=\cdots=\boldsymbol{y}(7)=\boldsymbol{u}(1)=\cdots=\boldsymbol{u}(7)$. We choose the controller orders with $L_y=n_y+1=2$, $L_u=n_u+1=4$ and controller parameters with $N$=4, $\boldsymbol{Q}=\boldsymbol{I}$ and $\boldsymbol{\lambda}=\boldsymbol{0}$.

The elements in PJM are calculated by $\hat{\boldsymbol{\Phi}}_1(k)=\hat{\boldsymbol{\Phi}}_3(k)=\hat{\boldsymbol{\Phi}}_4(k)=\boldsymbol{0}$,

$$
\hat{\boldsymbol{\Phi}}_2(k)=\begin{bmatrix}\hat{\phi}_{13}(k) & \hat{\phi}_{14}(k)\\ \hat{\phi}_{23}(k) & \hat{\phi}_{24}(k)\end{bmatrix}
=\begin{bmatrix}2y_1(k-2)+\Delta y_1(k-1) & 0.7(2y_2(k-2)+\Delta y_2(k-1))\\ 0.5(2y_1(k-2)+\Delta y_1(k-1)) & 1.3(2y_2(k-2)+\Delta y_2(k-1))\end{bmatrix},
$$

$$
\hat{\boldsymbol{\Phi}}_5(k)=\begin{bmatrix}\hat{\phi}_{1,9}(k) & \hat{\phi}_{1,10}(k)\\ \hat{\phi}_{2,9}(k) & \hat{\phi}_{2,10}(k)\end{bmatrix}=\begin{bmatrix}1 & 0.5\\ 0.4 & 1.2\end{bmatrix},
$$

$$
\hat{\boldsymbol{\Phi}}_6(k)=\begin{bmatrix}\hat{\phi}_{1,11}(k) & \hat{\phi}_{1,12}(k)\\ \hat{\phi}_{2,11}(k) & \hat{\phi}_{2,12}(k)\end{bmatrix},\ \hat{\phi}_{1,12}(k)=0.5,\ \hat{\phi}_{2,11}(k)=0.2,
$$

$$
\hat{\phi}_{1,11}(k)=0.4(u_1^2(k-4)+\frac{1}{2}\bullet 6u_1(k-4)\Delta u_1(k-3)+\frac{1}{6}\bullet 6\Delta u_1^2(k-3))
$$

$$
\hat{\phi}_{2,12}(k)=0.4(u_2^2(k-4)+\frac{1}{2}\bullet 6u_2(k-4)\Delta u_2(k-3)+\frac{1}{6}\bullet 6\Delta u_2^2(k-3))
$$

Similarly we can calculate $\hat{\boldsymbol{\Phi}}_1(k+1)$, $\cdots$, $\hat{\boldsymbol{\Phi}}_6(k+1)$, $\hat{\boldsymbol{\Phi}}_1(k+2)$, $\cdots$, $\hat{\boldsymbol{\Phi}}_6(k+3)$. We design the unconstrained implicit MPC (uiMPC) by minimizing the cost function (35) to have

$$
\Delta\boldsymbol{u}_{uiMPC}=\boldsymbol{g}^T\arg\min_{\Delta\boldsymbol{U}_N(k)} J \tag{58}
$$

and design constrained implicit MPC (ciMPC) by minimizing the following constrained cost function:

$$
\begin{aligned}
&\Delta\boldsymbol{u}_{ciMPC}=\boldsymbol{g}^T\arg\min J\\
&s.t.\ \ -5\leq u_1(k+i)\leq 0.6\\
&\qquad \sum_{j=1}^{2}\sum_{i=0}^{N-1}u_j^2(k+i)\leq 1
\end{aligned}
\tag{59}
$$

Fig. 1 and Fig. 2 show the tracking performance of the system controlled by uiMPC and ciMPC. Fig. 3 shows the control inputs. Fig. 4 shows the elements in the calculated PJM of uiMPC and Fig. 5 shows those of ciMPC.

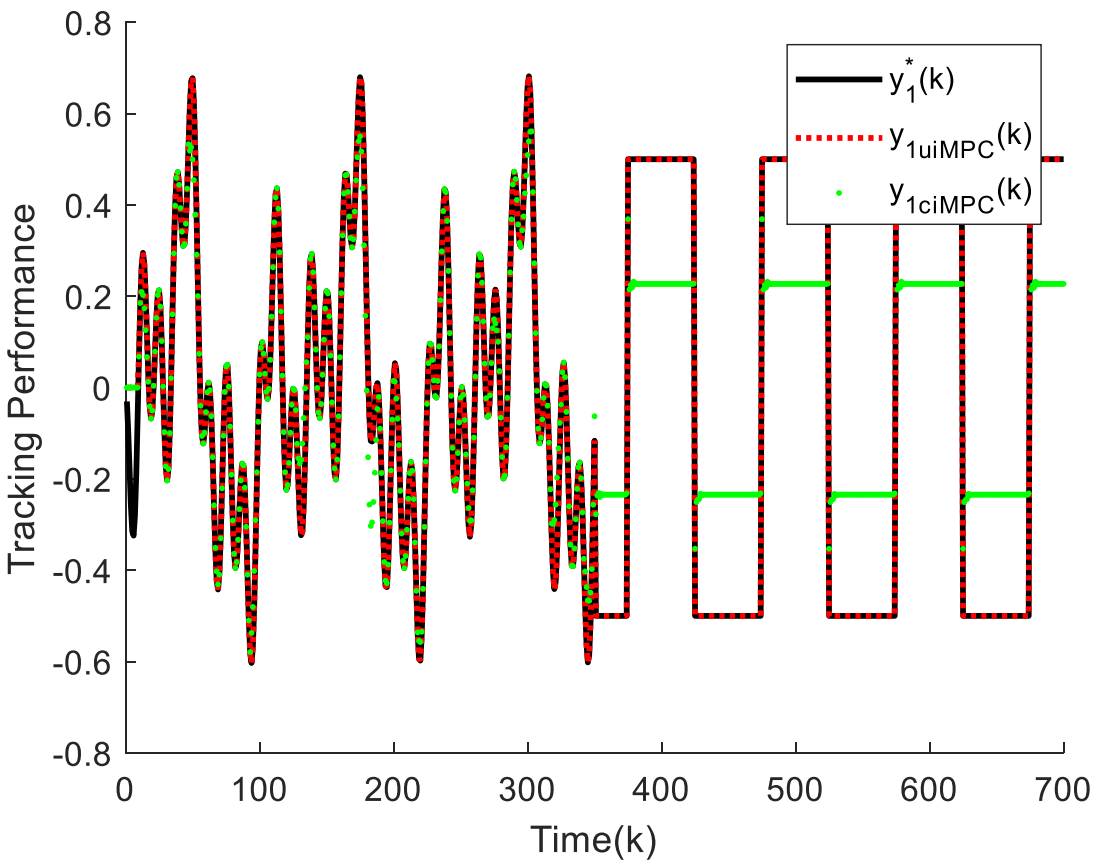

Fig. 1 Tracking performance $y_1(k)$

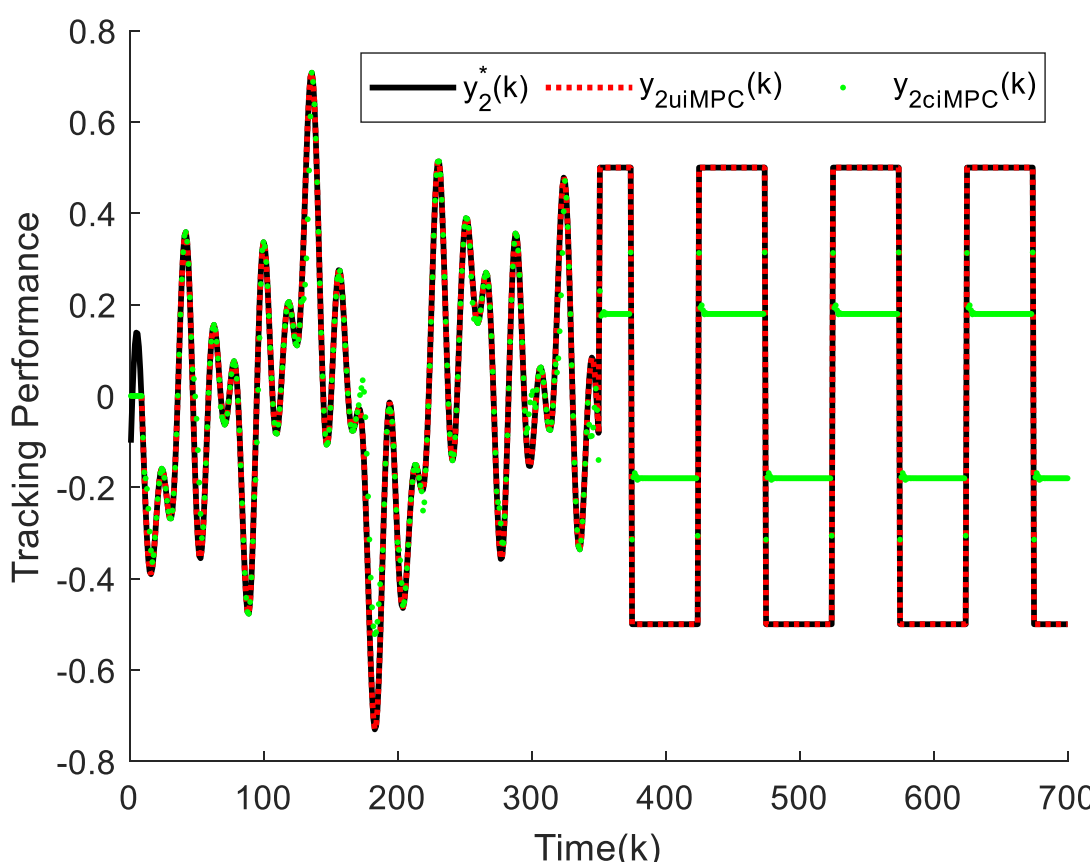

Fig. 2 Tracking performance $y_2(k)$

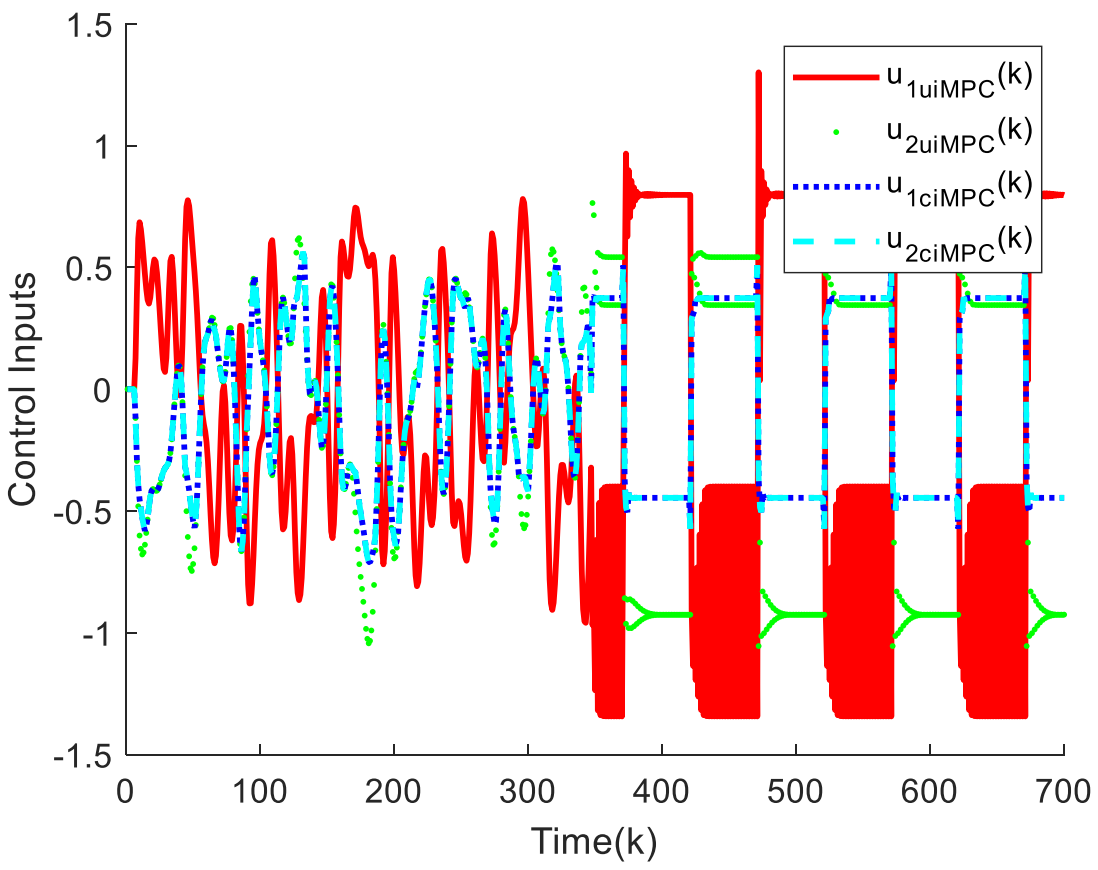

Fig. 3 Control inputs

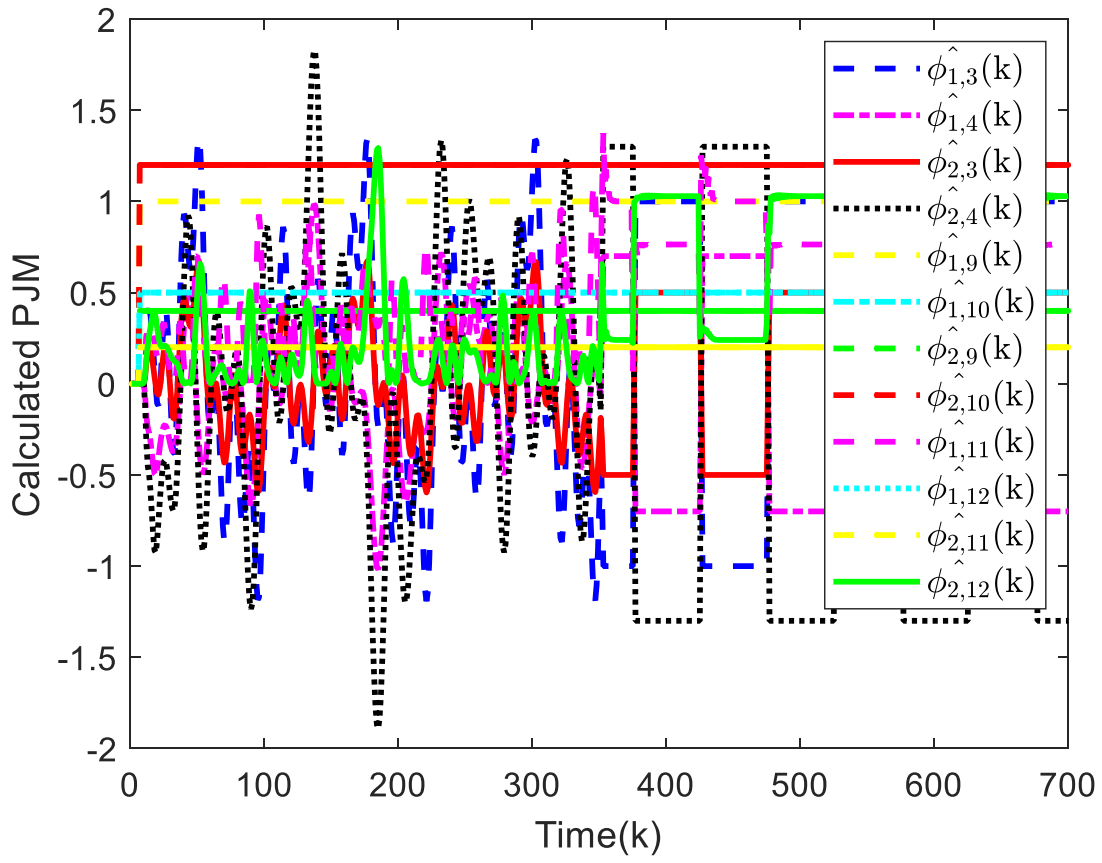

Fig. 4 Elements in calculated PJM of uiMPC

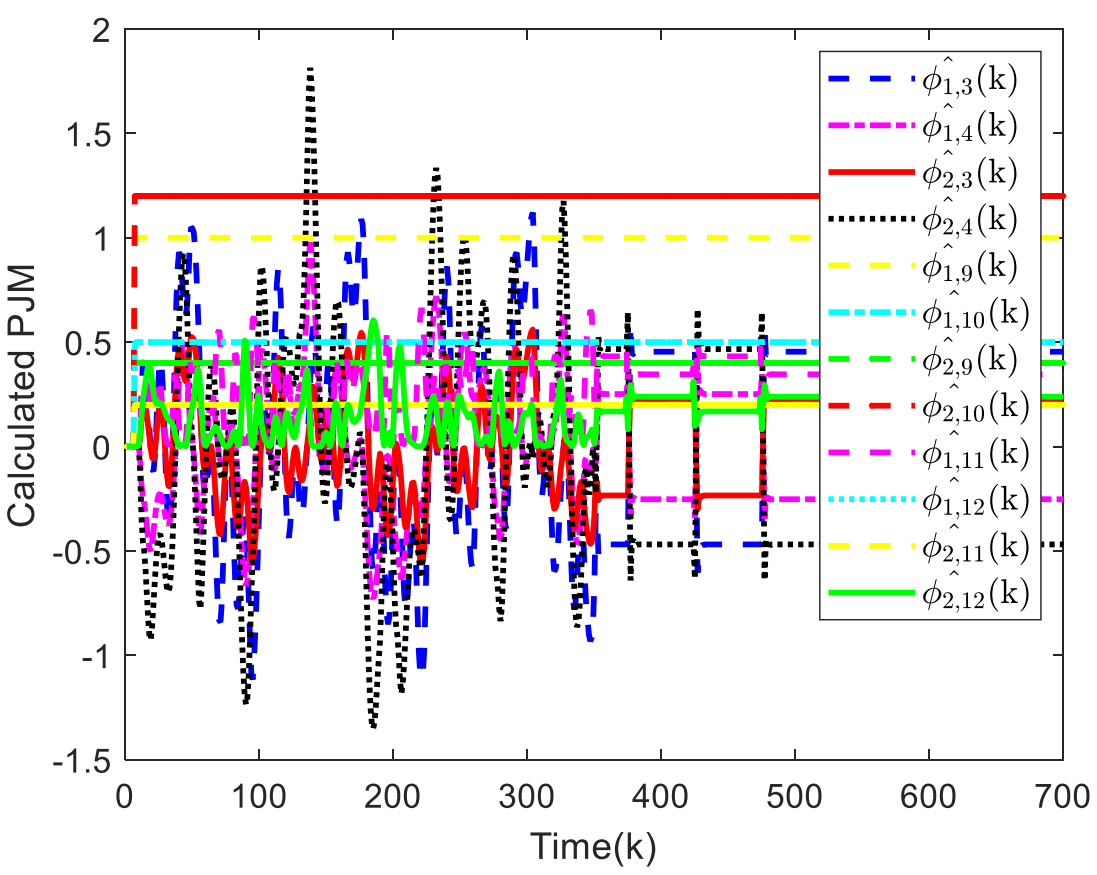

Fig. 5 Elements in calculated PJM of ciMPC

*Example 3*: Based on Example 2, we suppose that the known disturbance vector in model (56) is

$$\boldsymbol{w}(k+1)=\begin{bmatrix} w_1(k+1)\\ w_1(k+1)\end{bmatrix}=\begin{bmatrix}0.2\sin(k/10)+0.1\cos(k/30)\\ 0.1\sin(k/20)+0.2\cos(k/15)\end{bmatrix} \qquad (60)$$

All the controller settings are same as Example 2 except that we let $\Delta\hat{\boldsymbol{W}}(k+1)=\Delta\boldsymbol{W}(k+1)$. We design the unconstrained implicit MPC compensated with disturbance (uiMPC+D) by minimizing the cost function (35) to have

$$\Delta\boldsymbol{u}_{uiMPC+D}=\boldsymbol{g}^T \arg\min_{\Delta U_N(k)} J \qquad (61)$$

and design constrained implicit MPC compensated with disturbance (ciMPC+D) by minimizing the following constrained cost function:

$$\begin{aligned} &\Delta\boldsymbol{u}_{ciMPC+D}=\boldsymbol{g}^T \arg\min J\\ &s.t.\ \ -5\le u_1(k+i)\le 0.6\\ &\qquad \sum_{j=1}^{2}\sum_{i=0}^{N-1}u_j^2(k+i)\le 10 \end{aligned} \qquad (62)$$

Fig. 6 and Fig. 7 show the tracking performance of the system controlled by uiMPC+D and ciMPC+D. Fig. 8 shows the control inputs. Fig. 9 shows the elements in the calculated PJM of uiMPC+D and Fig. 10 shows those of ciMPC+D.

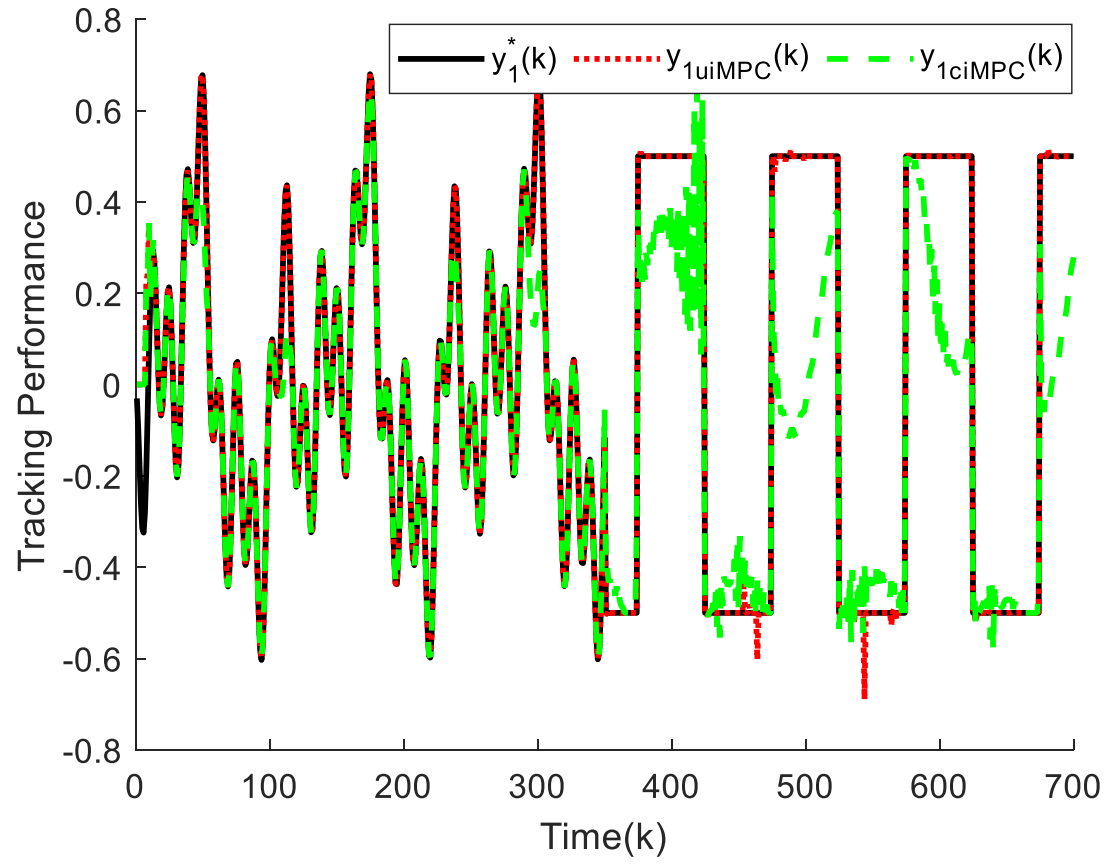

Fig. 6 Tracking performance $y_1(k)$

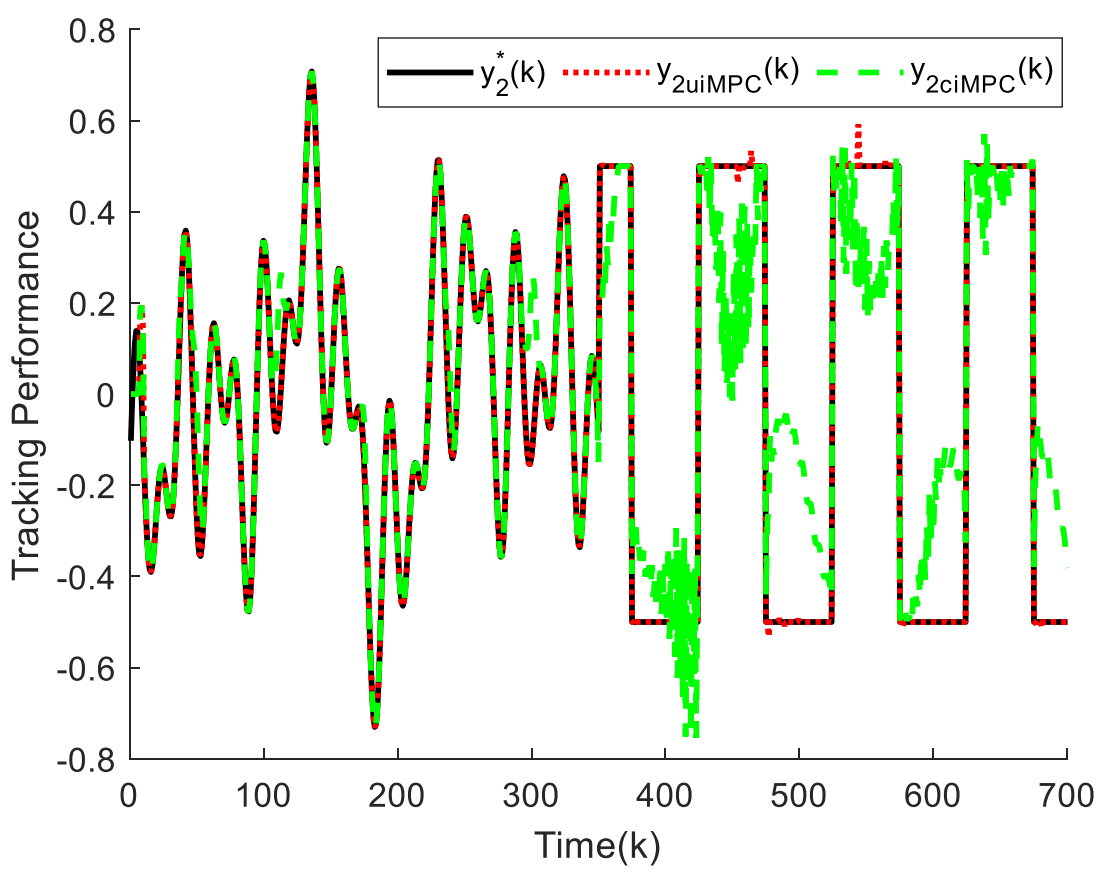

Fig. 7 Tracking performance $y_2(k)$

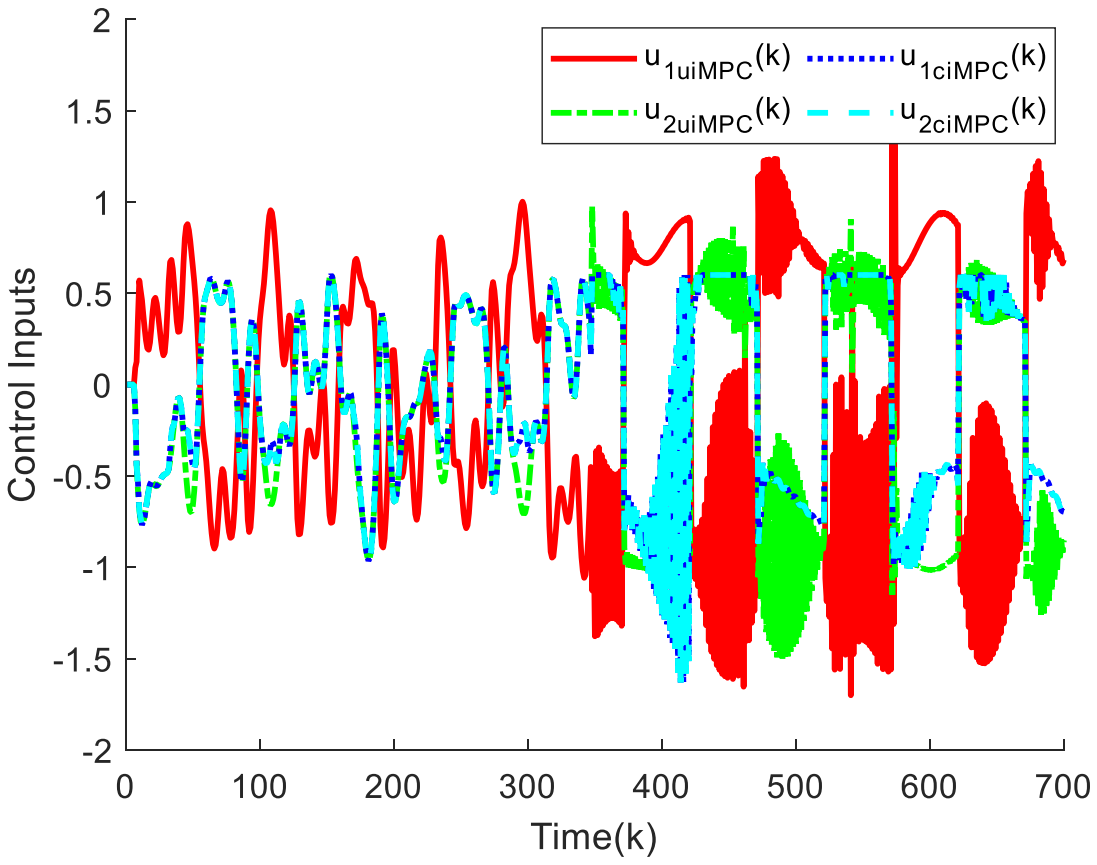

Fig. 8 Control inputs

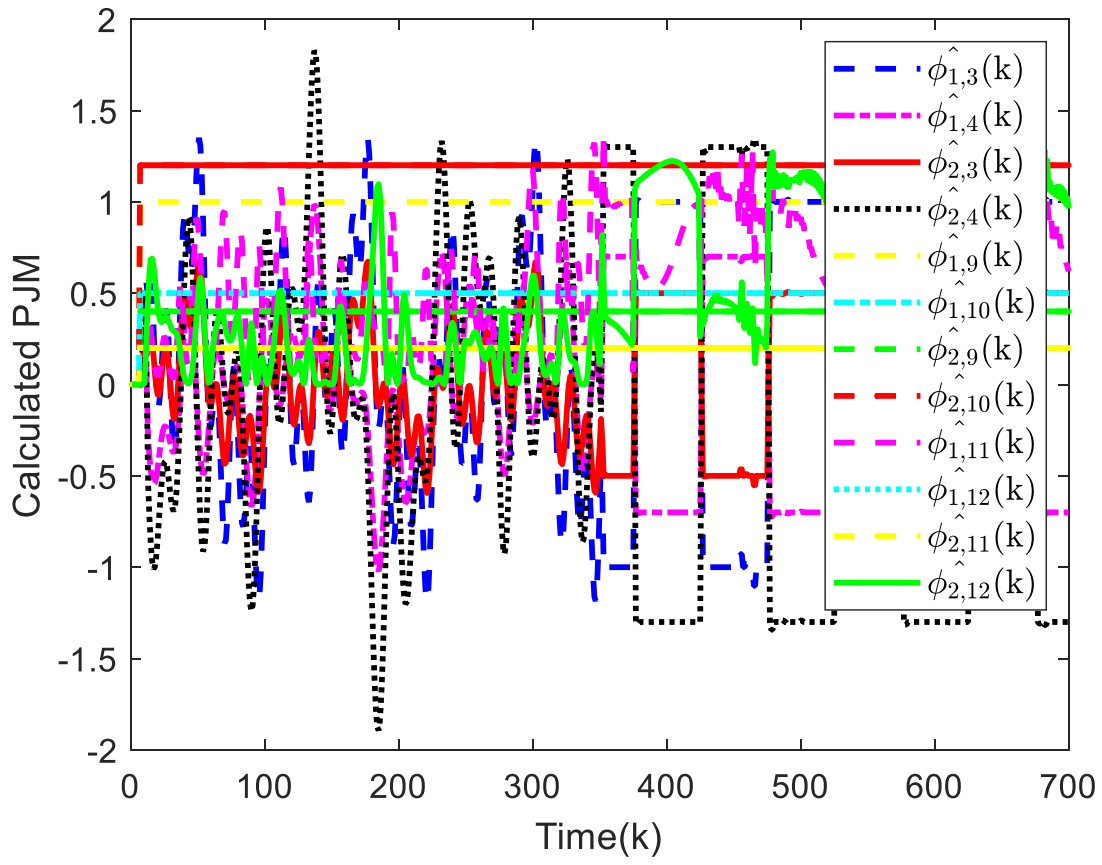

Fig. 9 Elements in calculated PJM of uiMPC

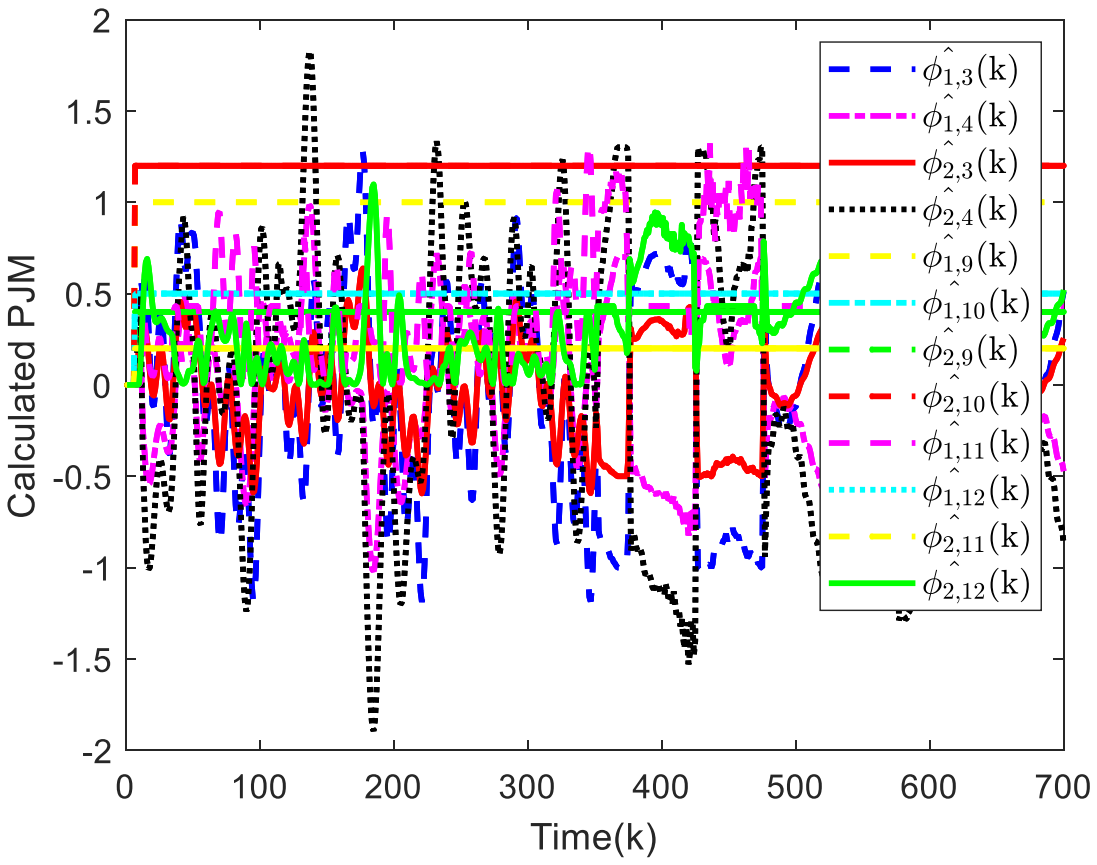

Fig. 10 Elements in calculated PJM of ciMPC

*Example 4*: We consider the following MIMO nonlinear system model:

$$
\begin{aligned}
y_1(k) &= y_1^2(k-2)+0.7y_2^2(k-2)+u_1(k-2)+0.5u_2(k-2) \\
&\quad +0.4u_1^3(k-3)+0.5u_2(k-3)+w_1(k) \\
y_2(k) &= 0.5y_1^2(k-2)+1.3y_2^2(k-2)+0.4u_1(k-2) \\
&\quad +1.2u_2(k-2)+0.2u_1(k-3)+0.4u_2^3(k-3)+w_2(k)
\end{aligned} \tag{63}
$$

where the unknown disturbance vector in model (63) is

$$
[w_1(k) \quad w_2(k)]^T = \begin{bmatrix} 0.3 & 0 \\ 0 & 0.2 \end{bmatrix} rand(2,1) \tag{64}
$$

where *rand*(2,1) is used to generate the white-noise sequences in Matlab.

We let $\Delta\hat{\boldsymbol{W}}(k+1)=\mathbf{0}$ to design the uiMPC in the same way as Example 2. We choose $N$=2, $\boldsymbol{Q}$=$\boldsymbol{I}$ and $\boldsymbol{\lambda}$=$\mathbf{0}$.

The closed-loop pulse transfer function for the disturbance is

$$
\frac{\boldsymbol{Z}(\boldsymbol{y}(k+1)}{\boldsymbol{Z}(\Delta\boldsymbol{w}(k+1))} = -\frac{\boldsymbol{Z}(\boldsymbol{y}^*(k+1)-\boldsymbol{y}(k+1))}{\boldsymbol{Z}(\Delta\boldsymbol{w}(k+1))} = (1+z^{-1})\boldsymbol{I} \tag{65}
$$

We define $[e_1(k),e_2(k)]^T$ as the measurement of the tracking error vector of the system and $[e_{d1}(k),e_{d2}(k)]^T=-\Delta\boldsymbol{w}(k)-\Delta\boldsymbol{w}(k-1)$ as the theoretically calculated tracking error vector due to the disturbances according to (65). Fig. 11 shows the tracking performance of the system controlled by uiMPC. Fig. 12 shows the elements in the calculated PJM of uiMPC. Fig. 13 and Fig. 14 show the contrasts between $e_i(k)$ and $e_{di}(k)$ through $e_i(k)/e_{di}(k)$ and $e_i(k)-e_{di}(k)$, ($i$=1, 2).

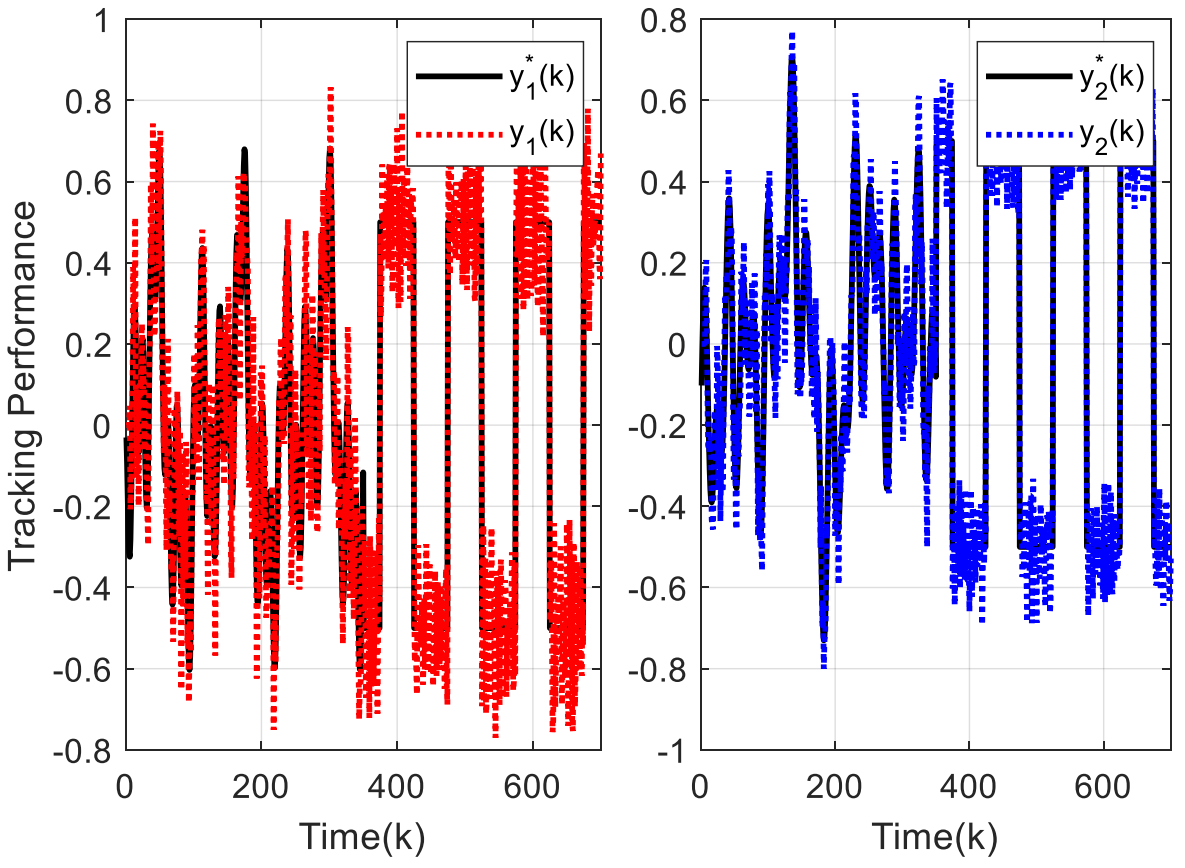

Fig. 11 Tracking performance

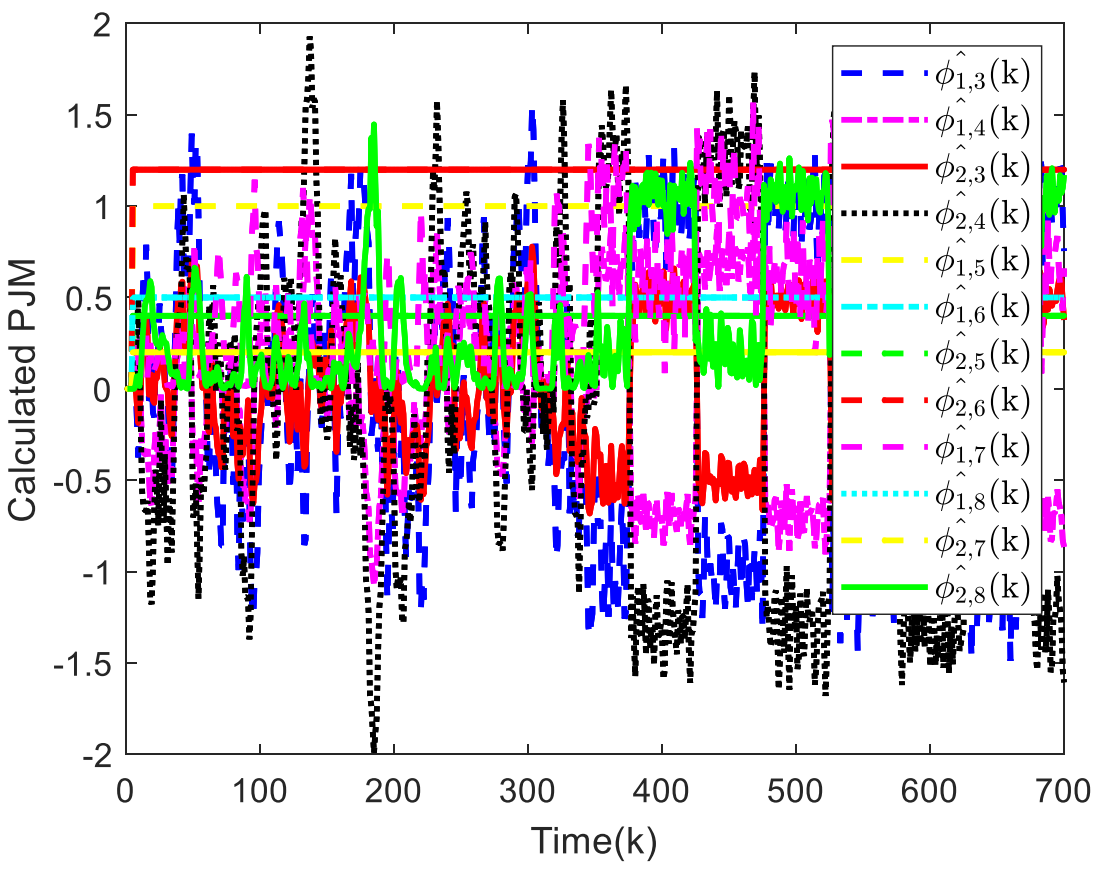

Fig. 12 Elements in calculated PJM of uiMPC

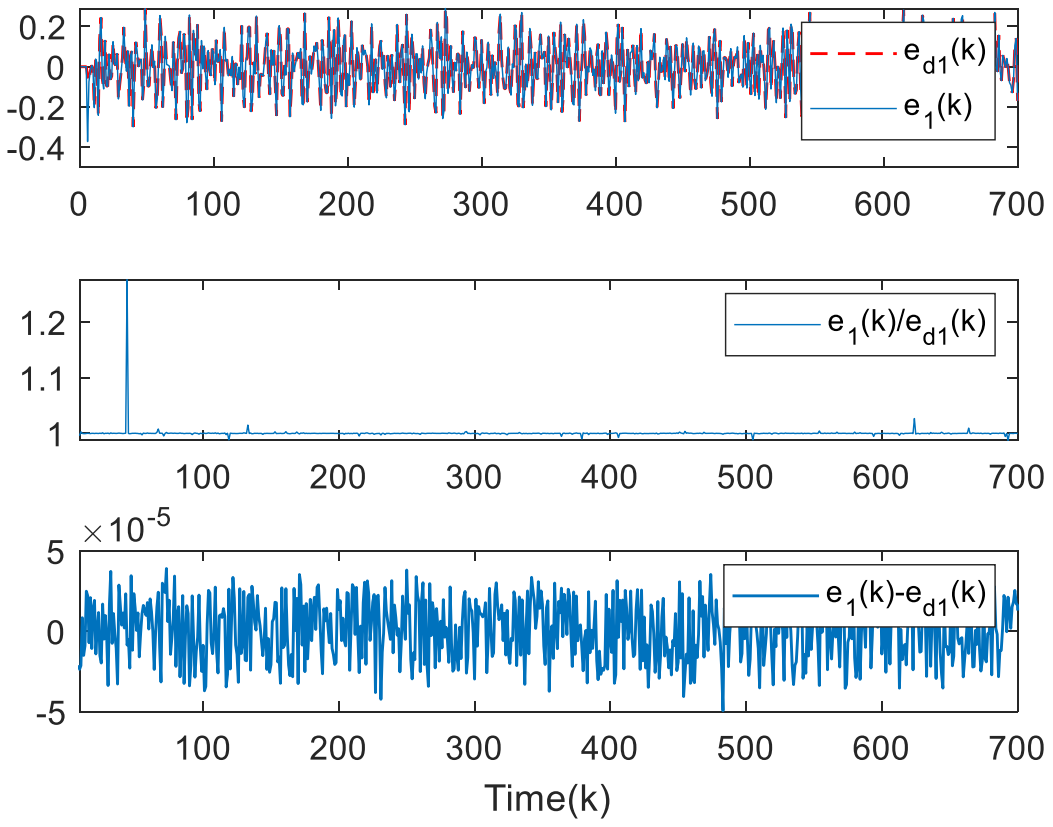

Fig. 13 Contrast between $e_1(k)$ and $e_{d1}(k)$

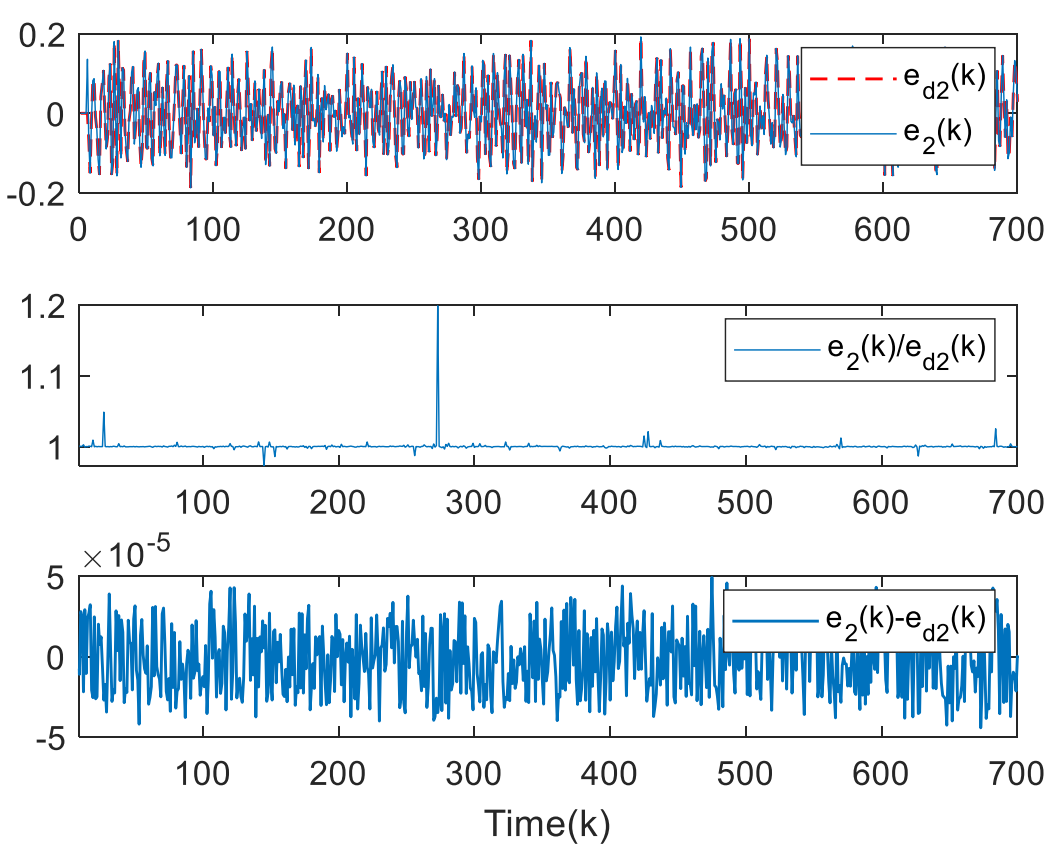

Fig. 14 Contrast between $e_2(k)$ and $e_{d1}(k)$

## IV. CONCLUSION

Based on EDLM, we propose a kind of MPC for SISO systems. After compensating the EDLM with disturbance for MIMO systems, the MPC compensated with disturbance is proposed to address the disturbance rejection problem. The system performances of both are analyzed by the closed-loop system characteristics, and the analyzed results are much clear compared with most system stability analyses on unconstrained MPC in current works. At last, simulations are carried out to validate the theories. One motivation of this paper is to promote the development of control theory and its application.

## APPENDIX

### Proof of *Theorem 1*

*Proof*: Case: $L_y=n_y+1$ and $L_u=n_u+1$

Based on the definition of the differentialbe function, (1) becomes

$$\begin{aligned}\Delta y(k+1) &= \frac{\partial f(\boldsymbol{\varphi}(k-1))}{\partial y(k-1)}\Delta y(k)+\cdots+\frac{\partial f(\boldsymbol{\varphi}(k-1))}{\partial y(k-n_y-1)}\Delta y(k-n_y)\\ &+\frac{\partial f(\boldsymbol{\varphi}(k-1))}{\partial u(k-1)}\Delta u(k)+\cdots+\frac{\partial f(\boldsymbol{\varphi}(k))}{\partial u(k-n_u-1)}\Delta u(k-n_u)\\ &+\gamma(k)\end{aligned} \tag{66}$$

where

$$\begin{aligned}\gamma(k) &= \varepsilon_1(k)\Delta y(k)+\cdots+\varepsilon_{Ly}(k)\Delta y(k-n_y)\\ &+\varepsilon_{Ly+1}(k)\Delta u(k)+\cdots+\varepsilon_{Ly+Lu}(k)\Delta u(k-n_u)\end{aligned} \tag{67}$$

We let

$$\begin{aligned}\boldsymbol{\phi}_L(k) = [&\frac{\partial f(\boldsymbol{\varphi}(k))}{\partial y(k)}+\varepsilon_1(k),\cdots,\frac{\partial f(\boldsymbol{\varphi}(k))}{\partial y(k-n_y)}+\varepsilon_{Ly}(k),\\ &\frac{\partial f(\boldsymbol{\varphi}(k))}{\partial u(k)}+\varepsilon_{Ly+1}(k),\cdots,\frac{\partial f(\boldsymbol{\varphi}(k))}{\partial u(k-n_u)}+\varepsilon_{Ly+Lu}(k)]^T\end{aligned} \tag{68}$$

to rewrite (66) as (3), with $(\varepsilon_1(k),\cdots,\varepsilon_{Ly+Lu}(k))\to(0,\cdots,0)$ in nonlinear systems, when $(\Delta y(k),\cdots,\Delta y(k-n_y),\Delta u(k),\cdots,$ $\Delta u(k-n_u))\to(0,\cdots,0)$. As to linear systems, we will always have $\boldsymbol{\phi}_L(k)=[\frac{\partial f}{\partial y(k)},\cdots,\frac{\partial f}{\partial y(k-n_y)},\frac{\partial f}{\partial u(k)},\cdots,\frac{\partial f}{\partial u(k-n_u)}]^T$, no matter what $(\Delta y(k),\cdots,\Delta y(k-n_y),\Delta u(k),\cdots,\Delta u(k-n_u))$ is.

Additionally, if the function $f(\cdots)$ has derivatives of all orders on any operating points, we can have (69) according to Taylor series

$$\begin{aligned}\Delta y(k+1) = [&\Delta y(k)\frac{\partial}{\partial y(k-1)}+\cdots+\Delta y(k-n_y)\frac{\partial}{\partial y(k-n_y-1)}\\ &+\Delta u(k)\frac{\partial}{\partial u(k-1)}+\cdots+\Delta u(k-n_u)\frac{\partial}{\partial u(k-n_u-1)}]f(\boldsymbol{\varphi}(k-1))\\ &+\frac{1}{2!}[\Delta y(k)\frac{\partial}{\partial y(k-1)}+\cdots+\Delta y(k-n_y)\frac{\partial}{\partial y(k-n_y-1)}\\ &+\Delta u(k)\frac{\partial}{\partial u(k-1)}+\cdots+\Delta u(k-n_u)\frac{\partial}{\partial u(k-n_u-1)}]^2 f(\boldsymbol{\varphi}(k-1))\\ &+\frac{1}{3!}[\Delta y(k)\frac{\partial}{\partial y(k-1)}+\cdots+\Delta y(k-n_y)\frac{\partial}{\partial y(k-n_y-1)}\\ &+\Delta u(k)\frac{\partial}{\partial u(k-1)}+\cdots+\Delta u(k-n_u)\frac{\partial}{\partial u(k-n_u-1)}]^3 f(\boldsymbol{\varphi}(k-1))\\ &+\cdots\end{aligned} \tag{69}$$

and then we can find a set of solution (70), (71) for (67) from (69).

$$\begin{aligned}\varepsilon_{i+1}(k) &= \frac{1}{2!}\frac{\partial^2 f(\boldsymbol{\varphi}(k-1))}{\partial y^2(k-i-1)}\Delta y(k-i)+\frac{1}{3!}\frac{\partial^3 f(\boldsymbol{\varphi}(k-1))}{\partial y^3(k-i-1)}\Delta y^2(k-i)\\ &+\frac{1}{4!}\frac{\partial^4 f(\boldsymbol{\varphi}(k-1))}{\partial y^4(k-i-1)}\Delta y^3(k-i)+\cdots\end{aligned} \tag{70}$$

$$\begin{aligned}\varepsilon_{Ly+1+j}(k) &= \frac{1}{2!}\frac{\partial^2 f(\boldsymbol{\varphi}(k-1))}{\partial u^2(k-j-1)}\Delta u(k-j)+\frac{1}{3!}\frac{\partial^3 f(\boldsymbol{\varphi}(k-1))}{\partial u^3(k-j-1)}\\ &\cdot\Delta u^2(k-j)+\frac{1}{4!}\frac{\partial^4 f(\boldsymbol{\varphi}(k-1))}{\partial u^4(k-j-1)}\Delta u^3(k-j)+\cdots\end{aligned} \tag{71}$$

, $i=0,\cdots,n_y$ and $j=0,\cdots,n_u$.